\documentclass[12pt,a4paper,twocolumn]{iopart}
\usepackage[T1]{fontenc}
\usepackage[latin1]{inputenc}
\usepackage{graphicx}
\usepackage{amssymb}
\usepackage{color}

\begin{document}

\title[]{Effect of the curvature and the $\beta$ parameter on the nonlinear dynamics of a drift tearing magnetic island}

\author{M. Muraglia$^{\rm 1,2,3}$, O. Agullo$^{\rm 1,2}$, M. Yagi$^{\rm 1,4}$, S. Benkadda$^{\rm 1,2}$, P. Beyer$^{\rm 1,2}$, X. Garbet$^{\rm 5}$, S.-I Itoh$^{\rm 1,4}$, K. Itoh$^{\rm 1,6}$ and A. Sen$^{\rm 7}$}

\address{\noindent$^{\rm 1}$France - Japan Magnetic Fusion Laboratory, LIA 336 CNRS\\
$^{\rm 2}$PIIM Laboratory, UMR 6633 CNRS-University of Provence, Marseille\\
$^{\rm 3}$Interdisciplinary Graduate School of Engineering Sciences, Kyushu University\\
$^{\rm 4}$RIAM, Kyushu University, Japan\\
$^{\rm 5}$CEA, IRFM, 13108, St-Paul-Lez-Durance, France\\
$^{\rm 6}$National Institute for Fusion Science, Japan\\
$^{\rm 7}$Institute for Plasma Research, Bhat, Gandhinagar 382428, India}
\begin{abstract}
We present numerical simulation studies of 2D reduced MHD equations investigating the impact of the electronic $\beta$ parameter and of curvature effects on the nonlinear evolution of drift tearing islands.
We observe a bifurcation phenomenon that leads to an amplification of the pressure energy, the generation of $E \times B$ poloidal flow and a nonlinear diamagnetic drift that affects the rotation of the magnetic island. These dynamical modifications arise due to quasi linear effects that generate a zonal flow at the onset point of the bifurcation.  Our simulations show that the transition point is influenced by the $\beta$ parameter such that the pressure gradient through a curvature effect  strongly stabilizes the transition. Regarding the modified rotation of the island, a model for the frequency is derived in order to study its origin and the effect of the $\beta$ parameter. It appears that after the transition, an $E \times B$ poloidal flow as well as a nonlinear diamagnetic drift are generated due to an amplification of the stresses by pressure effects.\\











\end{abstract}

\maketitle

\section{Introduction}

In tokamak and space plasmas, confinement can be affected by instabilities and in particular, at resonant surfaces, magnetohydrodynamics activity can lead to the generation of magnetic islands reaching a macroscopic width. Solar flares \cite{Priest81}, energy release events in the geotail \cite{Horton07} or tokamak internal disruptions, also known as sawtooth oscillations, are linked to such reconnection phenomena. Diamagnetic effects and self-generated zonal flows can modify the saturated island width via bifurcation mechanisms \cite{Ottaviani04}.  The rotation frequency of the island can also be nonlinearly affected with a strong dependence on  the transport coefficients and on the competition between the Reynolds and Maxwell stresses \cite{Nishimura08}. This can have a significant physical consequence, for example in a tokamak, where such a nonlinear effect on the rotation can lead to a slowing down of the plasma through locking to the resistive wall producing in turn a degradation of the plasma and/or triggering a transport barrier \cite{Itoh06}.
Likewise curvature effects can also modify the nature of island dynamics. Magnetic islands can in particular coexist with pressure driven intabilities such as interchange modes and/or turbulence. Several experiments report the coexistence of turbulence and MHD activities showing some correlated effects \cite{Tanaka05, Takaji02}. Numerical studies of the interaction between double tearing modes and micro-turbulence to
delineate the interaction between zonal flows and the latter in the growing phase of the double tearing instability
have also been performed in  \cite{Ishizawa07}. More recently, in \cite{Militello08} an investigation of the interaction of a 2D electrostatic
turbulence with an island whose dynamics is governed by a generalized Rutherford
equation has been carried out.  However the study neglects the potentially stabilizing influence of the magnetic structure on the turbulence precluding thereby any multi-scale interaction between MHD and turbulence. In this paper we study the dynamics of a magnetic islands in the presence of interchange effects but limit ourselves to the situation where the interchange modes in the system are linearly stable. We find that the whole system does not generate turbulence in the nonlinear stage but exhibits a complex dynamics arising mainly due to quasilinear effects. 
Our investigations are based on linear and nonlinear simulations of a set of reduced fluid equations (a three field model) through which we examine the origin and the influence of zonal flows on the magnetic island dynamics in the presence of interchange effects. The magnitude of the pressure gradient appears to be a key parameter of the dynamics controlling both, the generation of the zonal flow and the development of a nonlinear transition in the system. \\
The paper is organized as follows. In section II, the model equations are introduced. In section III, a linear analysis of the model is done in order to understand the role of the equilibrium magnetic field in the stabilization of the electromagnetic interchange modes. In section IV, the description and the analysis of the dynamics are done. In section V, the origin of the island poloidal rotation is investigated. Section VI presents a summary and conclusions of the paper.

\section{Model system}
Our model system is a three fields model corresponding to a reduced magnetohydrodynamic description of the fluid equations \cite{Hazeltine} and which provides a minimal framework for including both the interchange and the tearing 
mode phenomena in a plasma. The model consists of a set of three coupled equations for the electrostatic potential $\phi$, the pressure of the electron $p$ and the magnetic flux $\psi$. We suppose that the magnetic field is dominated by a constant component $B_{0z}$ along the z-direction. The time evolution of the three fields are described by:
\begin{eqnarray}
\hskip-1.5cm
\partial_{t}\nabla_{\perp}^{2}\phi+\left[\phi,\nabla_{\perp}^{2}\phi\right]&=&
\left[\psi,\nabla_{\perp}^{2}\psi\right]-\kappa_{1}\partial_{y}p+\nu\nabla_{\perp}^{4}\phi,\label{eq:equPHI} \\
\hskip-1.5cm
\partial_{t}p+\left[\phi,p\right]&=&
  -v_{\star}\biggl((1-\kappa_{2})\partial_{y}\phi+\kappa_{2}\partial_{y} p\biggr) +C^2\left[\psi,\nabla_{\perp}^{2}\psi\right]+\chi_{\perp}\nabla_{\perp}^{2}p,\label{eq:equPE}\\
%
\hskip-1.5cm
 \partial_{t}\psi+\left[\phi-p,\psi\right]&=&-v_{\star}\partial_{y}\psi+\eta\nabla_{\perp}^{2}\psi,\label{eq:equPSI} 
 \end{eqnarray}

 \noindent where $v_{\star}=\frac{\beta L_{\perp}}{2\Omega_{i}\tau_{A}L_{p}}$. The
sum of the electron and ion momentum evolution equations leads to the
plasma equation of motion, Eq. (\ref{eq:equPHI}), where $\nu$ is the
viscosity. Eq. (\ref{eq:equPE}) comes from the energy conservation equation where
$\chi_{\perp}$ is the diffusivity. Eq. (\ref{eq:equPSI}) is 
Ohm's law (electron parallel momentum equation) with $\eta$ being the resistivity.
$\beta=\frac{p_{0}}{B_{0z}^{2}/2\mu_{0}}$ is the ratio of the electron thermal energy to the
magnetic energy ($p_{0}$ being the amplitude of the equilibrium pressure), $L_{p}$ is the pressure gradient length,
$L_{\perp}$ is a magnetic shear length,
$R_{0}$ is the major plasma radius, $\Omega_{i}=\frac{eB_{0z}}{m_{i}}$ is
the ion cyclotron frequency,  and $\tau_{a}$ is the Alfv\`en time.
Equations (\ref{eq:equPHI}-\ref{eq:equPSI}) are
normalized as follow:
\begin{equation}
\frac{t}{\tau_{A}}\rightarrow t,\quad
\frac{x}{L_{\perp}}\rightarrow x,
\end{equation}
\begin{equation}
\frac{\psi}{L_{\perp}B_{0z}}\rightarrow \psi,\quad
\frac{\phi}{L_{\perp}v_{A}B_{0z}}\rightarrow\phi,\quad
\frac{L_p}{L_{\perp}p_{0}}p\rightarrow p,
\end{equation}
where $v_{A}=B_{0z}/\mu_{0}nm_{i}=L_{\perp}/\tau_{a}$ is the characteristic Alfv\`en speed.
\noindent $\kappa_{i}$ parameters are linked to the curvature and to the
pressure gradient
($\kappa_{1}=2\Omega_{i}\tau_{A}\frac{L_{\perp}}{R_{0}}$ and
$\kappa_{2}=\frac{10L_{p}}{3R_{0}}$), so these parameters control the
interchange instability. On the other hand, in Eq. (\ref{eq:equPE}),
the tearing mode dynamics is controlled by the coupling parameter
$C^2=\frac{5\beta}{6\Omega_i^2\tau_A^2}$.  More precisely,
this parameter controls the coupling between pressure and the
magnetic flux. The nature of the linear and nonlinear dynamics of the
magnetic island depends strongly on the strength of the
coupling. For a high $\beta$ plasma, since the coupling is strong, the pressure and the magnetic flux control the island dynamics,
whereas for a low $\beta$ plasma, the island dynamics is governed by the interaction between the flow and the magnetic
flux. In our model we assume the electron temperature to be constant and the ions to be cold. The cold ion limit is physically realistic since the ion temperature does not significantly affect
the stability of the tearing mode. As a further simplification we have also neglected the parallel ion
dynamics in the energy balance equation Eq.
(\ref{eq:equPE}).
Eqs.~(\ref{eq:equPHI}--\ref{eq:equPSI}) are solved numerically using a finite difference scheme in the $x$ 
direction, including an Arakawa algorithm \cite{Arakawa97} for an accurate conservation of the Poisson brackets $[.,.]$ and a pseudo-spectral method in the $y$ direction, including an appropriate de-aliasing scheme.

\section{Nature of the tearing modes and the influence of the curvature parameter $\kappa_1$}

We now study the influence of the interchange mechanism on the magnetic reconnection when the gradient scale length of the pressure, $L_p$, is of the order of the size of the island: we set $L_p=L_{\perp}$. We are interested in large islands, i.e islands with widths $w$ such that $a \gtrsim w \gg \rho_s$ where $a$ is the minor radius and $\rho_s$ is the hybrid Larmor radius ($\rho_s = c_s/\Omega_i$, where $c_s$ is the ion sound velocity), and  we have chosen $L_{\perp}=0.24$m.  The numerical values of other parameters are taken to be $R_0=2.24$m  and $\Omega_{i}\tau_{A}$=0.5.  These numerical values are typical of the TORE SUPRA device for an island width of about $1/3$ the minor radius and lead to $\kappa_1\sim 0.11$ and $\kappa_2\sim 0.36$.  The widths of the numerical integration box are set to $L_x=2\pi L_{\perp}$ and $L_y=5\pi L_{\perp}$. The values of the coefficients $C(\beta)$ and $\omega_\star(\beta)=k_yv_\star$ are determined for four different values of $\beta$  in the range of $10^{-3}$ to $2.5 \times 10^{-2}$ (see  Table (\ref{fig:tableau})).  
\begin{table}
\begin{center}
\begin{tabular}{|c|c|c|}
\hline
$\beta$   &  $v^\star$ & $C^2$\\ \hline
$0.001$   &  $2*10^{-3}$ & $3.3*10^{-3}$\\ \hline
$0.005$   &  $10^{-2}$ & $1.67*10^{-2}$\\ \hline
$0.015$   &  $3*10^{-2}$ & $5*10^{-2}$\\ \hline
$0.025$   &  $5*10^{-2}$ & $8.33*10^{-2}$\\ \hline
\hline
\end{tabular}
\end{center}
\caption{Effect of the $\beta$ parameter on $v^\star$ and $C$.}
\label{fig:tableau}
\end{table}
Transport coefficients $(\nu,\eta,\chi)$ are all set to $10^{-4}$ which correspond to renormalized coefficients to include effects of microscopic turbulence \cite{Furuya02} .
The equilibrium magnetic field ${\bf B_{0y}}=B_{0}\hat{{\bf y}}$, based on the Harris current sheet model~\cite{Harris62}, is chosen to be
of the form,
\begin{equation}
 B_{0}(x)=\tanh \left( \frac{x - L_x/2}{a_t}\right).
 \label{eq:B0}
\end{equation}
The parameter $a_t=0.75$ controls the width of the profile, $\psi_0'(x)=B_0(x)$. With such a profile, the parameter $\Delta^{\prime}$ (the tearing mode stability index) can be explicitly computed taking into account the boundary conditions, for modes evolving slowly on the Alfv\'en time scale and further neglecting the viscous and interchange corrections. Introducing $\hat k=a_t k_y$, we have

\begin{equation}
a_t\Delta^{\prime} = 2 \left( 1/\hat k- \hat k \right)+\hat\Delta_{\mbox{b.c}}\;,
\end{equation}
where 
\begin{equation}
2\hat\Delta_{\mbox{b.c}}^{-1}=-\int_0^{L_x/(2a_t)} dy \exp(2\hat ky)/(1+\tanh(y)/\hat k)^ 2 
\end{equation}
is a correction linked to the finite radial distance of the walls.\\

\begin{figure}
\begin{center}
\includegraphics[width=7cm,height=5cm]{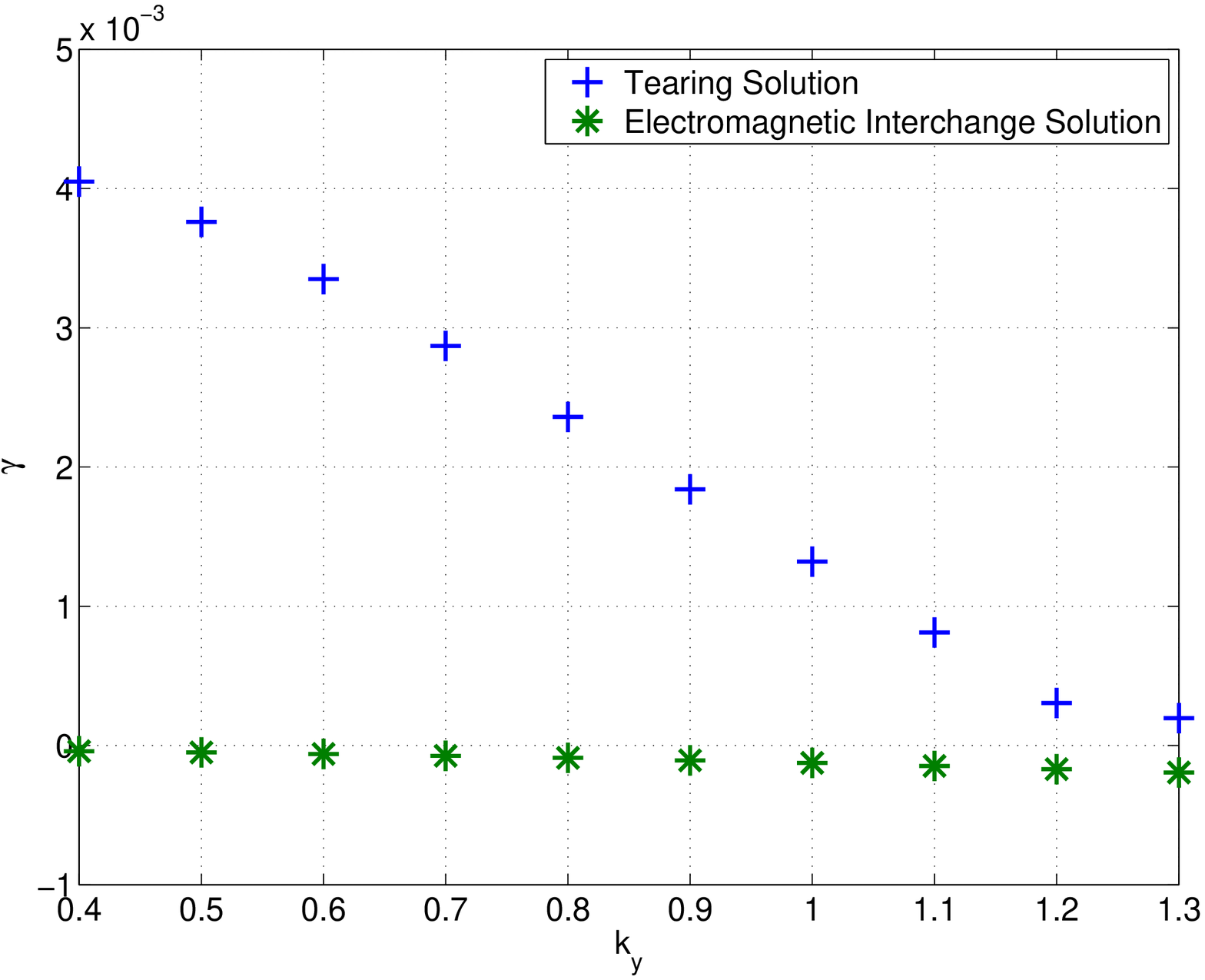}
\includegraphics[width=7cm,height=5cm]{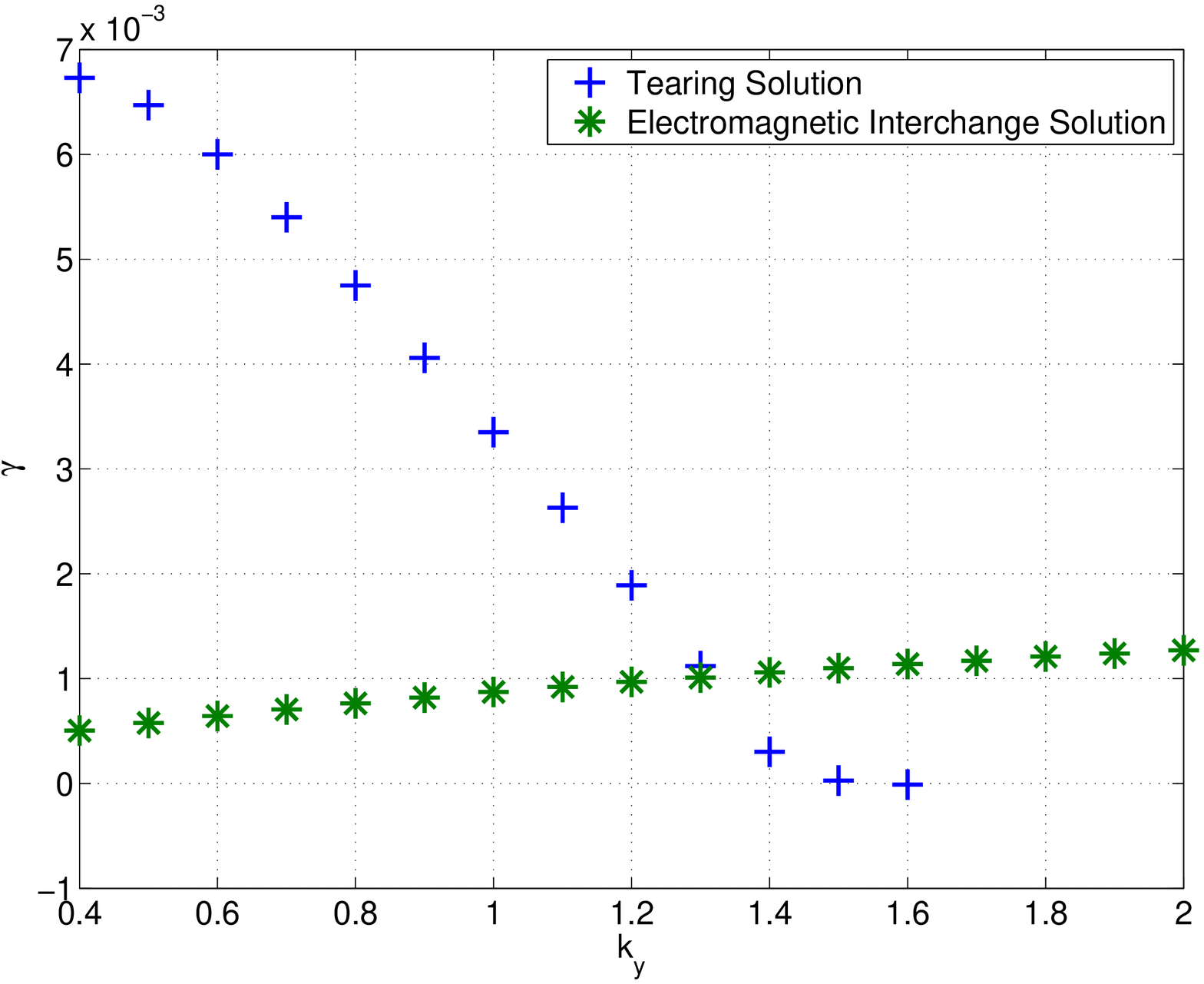}
\end{center}
\caption{Linear growth rate $\gamma$ $[\tau_A^{-1}]$ versus poloidal mode number $k_y$ for a simulation with $\beta = 10^{-3}$.  [Left]  $v^\star=2.10^{-3}$, $C=3.33.10^{-3}$, $\kappa_1=0.1071$, $\kappa_2 = 0.3571$ and $\mu=\chi_\perp=\eta=10^{-4}$. [Right]  $v^\star=10^{-2}$, $C=2.10^{-3}$, $\kappa_1=5$, $\kappa_2 = 0.3571$, $\mu=\chi_\perp=10^{-5}$ and $\eta=10^{-4}$.}
 \label{fig:figure3_b}
\end{figure}

Let us investigate the stability of the modes modeled by eqs.~(\ref{eq:equPHI}--\ref{eq:equPSI})  with the given numerical values of the parameters in the presence of such an equilibrium.
 Figure (\ref{fig:figure3_b}), shows the growth rate of the electromagnetic interchange and the tearing branches, as functions of the poloidal wave number, for the parameter values given above. The left graph has been obtained using the parameters chosen in this work with $\beta=0.001$. Tearing instability has the largest growth rate at $k_y^{\mbox{\tiny tear}}=2\pi/LY=0.4$, for which $\Delta'=7$ and  $\gamma_{\mbox{\tiny tear}}\sim0.0042$. This is clearly smaller than the one we would obtain in the classical tearing limit , i.e if we would have set all the parameters to zero except $\eta$ ($\gamma_{\mbox{\tiny tear class}}= 0.0072$). 
It is instructive to note also that in this parametric regime the interchange branch is stable for any wave number. From an electrostatic point of view, with such parameters, interchange would have been unstable for $k_y<8$ and would have given a scale separation between both instabilities ($k_y^{\mbox{\tiny int elec}}/k_y^{\mbox{\tiny tear}}\sim 7$, $\gamma^{\mbox{\tiny int elec}}/\gamma^{\mbox{\tiny tear}}\ll 1$). Let us focus on the tearing branch. Linearisation of eqs.~(\ref{eq:equPHI}--\ref{eq:equPSI}) in the vicinity of the resonnance shows that curvature effects weakly modify  the growth rate if $\kappa_1 v_\star(k_y/k_x)^2/\gamma^2\ll1$ and $\kappa_2\ll 1$, which is true in our case. Considering the fact that the linear regime is also not controlled by viscous phenomena \cite{Porcelli87}, it follows that, linearly, this system develops approximately drift tearing modes. The actual nature of these modes is controlled mainly by the ratio $P/P_{\mbox{cr}}$ and $\gamma/\omega_\star$, where $P=\nu/\eta=1$ is the Prandtl number and $P_{\mbox{cr}}=(\Delta'(\eta/k_y)^{1/3})^{6/5}$. In our cases, for any $\beta$, the first ratio is always smaller than $1$. It implies that when  $\gamma/\omega_\star>1$, one gets the visco-tearing regime with a growth rate scaling law $\gamma_{vt}\sim 0.47 \Delta'\eta^{2/3}P^{-1/6}k_y^{1/3}$.
When  $\gamma/\omega_\star<1$, we recover the visco-drift-tearing regime with the growth rate $\gamma_{vdt} $ \cite{Grasso01}. For instance, for $\beta=0.001$ and $k_y=0.4$, we have $\gamma_{vt}/\omega_\star\sim 6.5> 1$ and $\gamma=0.00042\sim\gamma_{vt}=0.00052$. 
For $k_y=1.2$, we have $\gamma_{vt}/\omega_\star\sim 0.2< 1$ and $\gamma=0.00035 \sim\gamma_{vdt}=0.0004$. 
%
 %
 %
%
%
%
%
%

The right graph of Figure (\ref{fig:figure3_b}) shows that there exist regimes where the interchange branch is unstable and has the largest growth rate at small scales. The study of such regimes is out of the scope of this paper. We also remark that the instability does not necessarily develop in the vicinity of a resonant surface, but in that case, the effect of the magnetic field on the stability of  interchange like modes  can be investigated by setting  $\psi_0'=B_0=\mbox{Cte}$ and using some Fourier analysis. 
%
%
%
\begin{figure}
\begin{center}
\includegraphics[width=8cm,height=5cm]{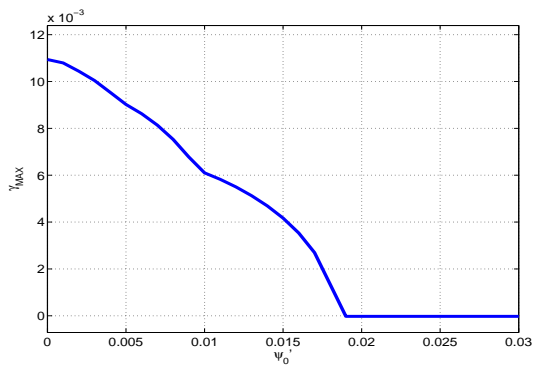}
\end{center}
\caption{Stability of electromagnetic interchange modes versus $\psi_0'$, far from the resonant surface. Same parameter values as figure 1.}
 \label{fig:figure3_a}
\end{figure}
Figure (\ref{fig:figure3_a}) shows the linear growth rate of interchange modes versus $\psi_0'$. As is well-known \cite{Biskamp0093}, the equilibrium magnetic field stabilises the interchange modes and, in our case $B_0=1$, this is clearly stable. 
We can therefore expect that in the initial phase the growth of the magnetic island is weakly influenced by interchange parameters.\\


\section{Nonlinear generation of a strong zonal flow}

\subsection{Description of the nonlinear evolution of the system}
\begin{figure}
\begin{center}
\includegraphics[width=10cm,height=7cm]{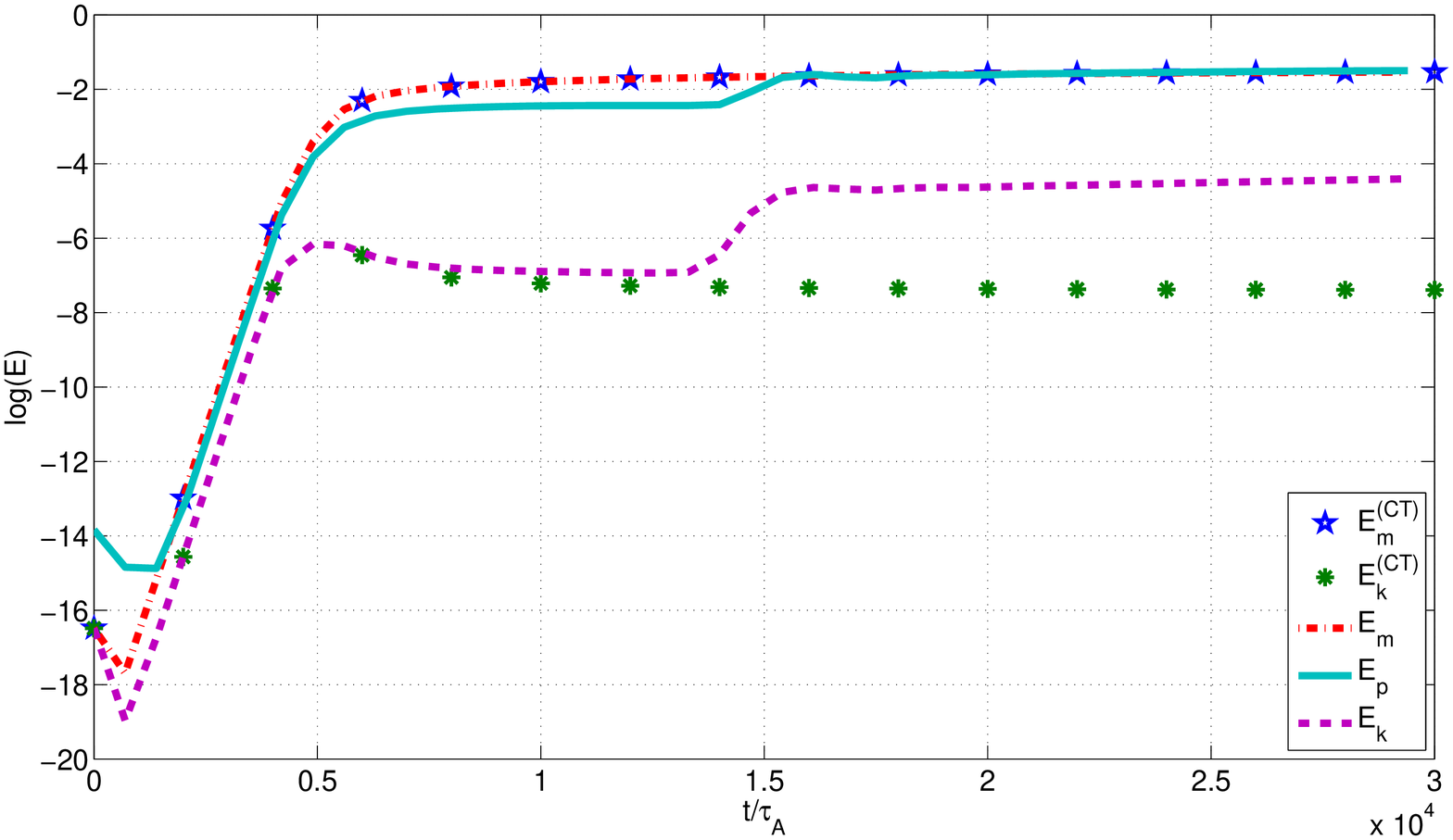}
\end{center}
\caption{Time Evolution of the Energies for $\beta = 10^{-3}$.}
 \label{fig:figure4_a}
\end{figure}
To characterize how the pressure gradient affects the evolution of a
magnetic island, linear and nonlinear self consistent numerical simulations  have been performed. A grid number of $n_{x}=128$ is chosen for
the radial direction and $n_{y}=128$ for the poloidal direction (equivalent to 48 modes in this direction, including dealiasing). 
The energy conservation relation derived from Eqs. (\ref{eq:equPHI}--\ref{eq:equPSI}) is
\begin{equation}
\frac{d}{dt}\left(E_m+E_p+E_k\right)=-\eta <j^2>-\nu <\Delta^2\phi>-\frac{\chi_\perp}{C^2}<\left|\nabla p\right|^2>+S\;,
\label{eq:energy}
\end{equation}
where $E_m = 0.5 <\!\!\left|\nabla(\psi-\psi_{0})\right|^2\!\!>$, $E_p= 0.5 <\!\!p^2\!\!>\!\!/C^2$ and $E_k= 0.5 <\!\!\left|\nabla\phi\right|^2\!\!>$ are respectively the magnetic energy, the pressure energy and the kinetic energy of the fluctuations. The brackets $<.>$ mean here an average over the simulation domain.
%
$S$ is the source term linked to the curvature and the pressure gradient, proportional to the radial pressure flux ,
%
%
$
S= -\alpha_S < p\partial_y\phi > 
$
%
with $\alpha_S= \frac{v^\star}{C^2}(1-\kappa_2)+\kappa_1 >0 $  because $\kappa_2<1$. Note that a local flattening of the pressure by radial exchange of pressure cells, gives a fluctuation $\delta S<0$.  Moreover, the interchange  source term $S$ is not modified by the generation of zonal flow. 
Figure (\ref{fig:figure4_a}) shows the time evolution of $E_m$, $E_p$ and $E_k$ for the parameters chosen in this work with $\beta=10^{-3}$
 as well as the corresponding $E_m^{CT}$ and $E_k^{CT}$, for a classical tearing mode (i.e $p=0$ and $\kappa_i=0$).
In comparison with the evolution of a classical tearing mode, four regimes are observed in the nonlinear simulations of a magnetic island in the presence of the interchange term. First, there is a linear regime where the magnetic island is formed. Second, the system reaches a quasi-plateau phase. Then, a transition occurs and as it will be shown later this is linked to the interchange parameters. Finally, the system reaches a new saturated state.
\begin{figure}
\begin{center}
\includegraphics[width=4.5cm,height=5cm]{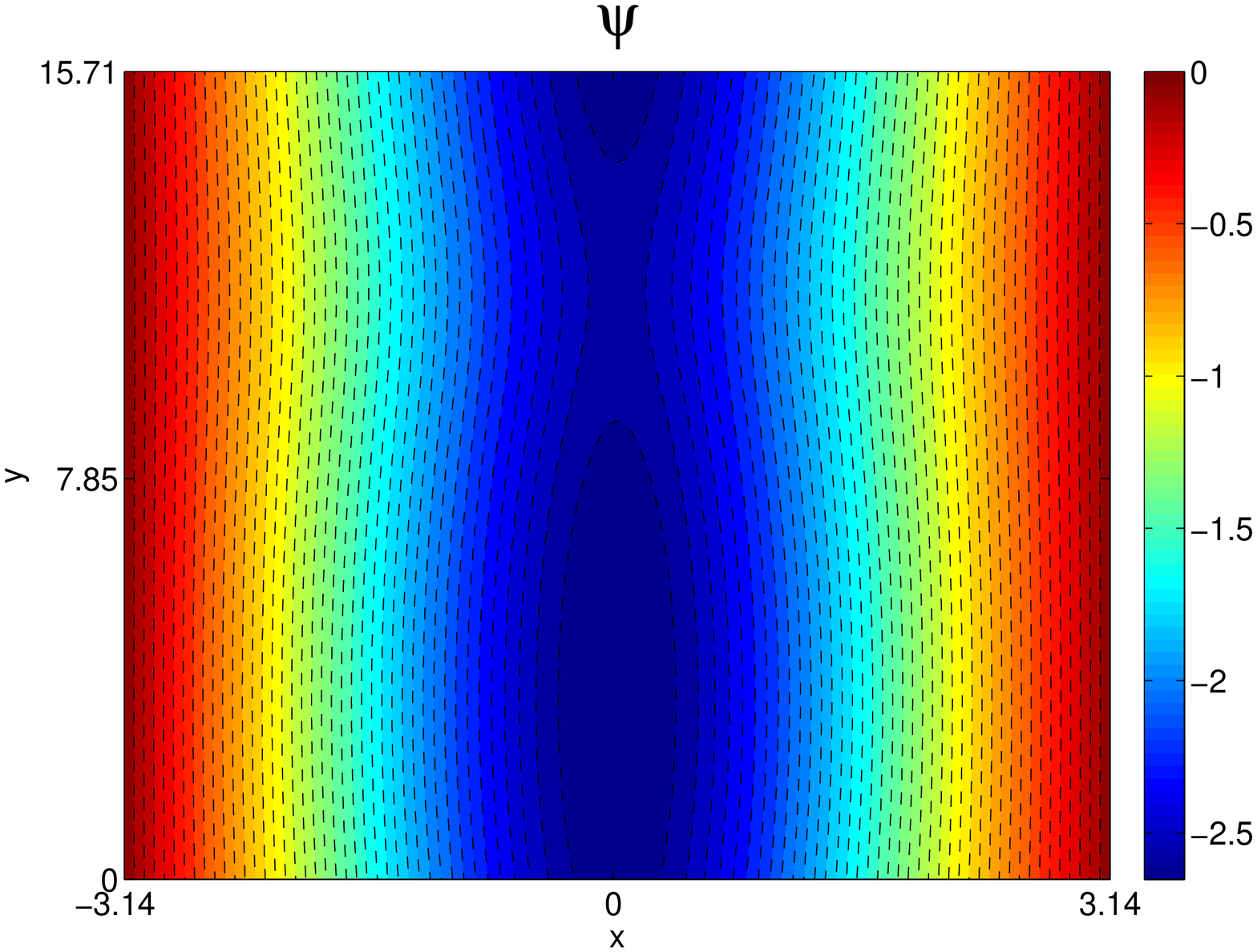}
\includegraphics[width=4.5cm,height=5cm]{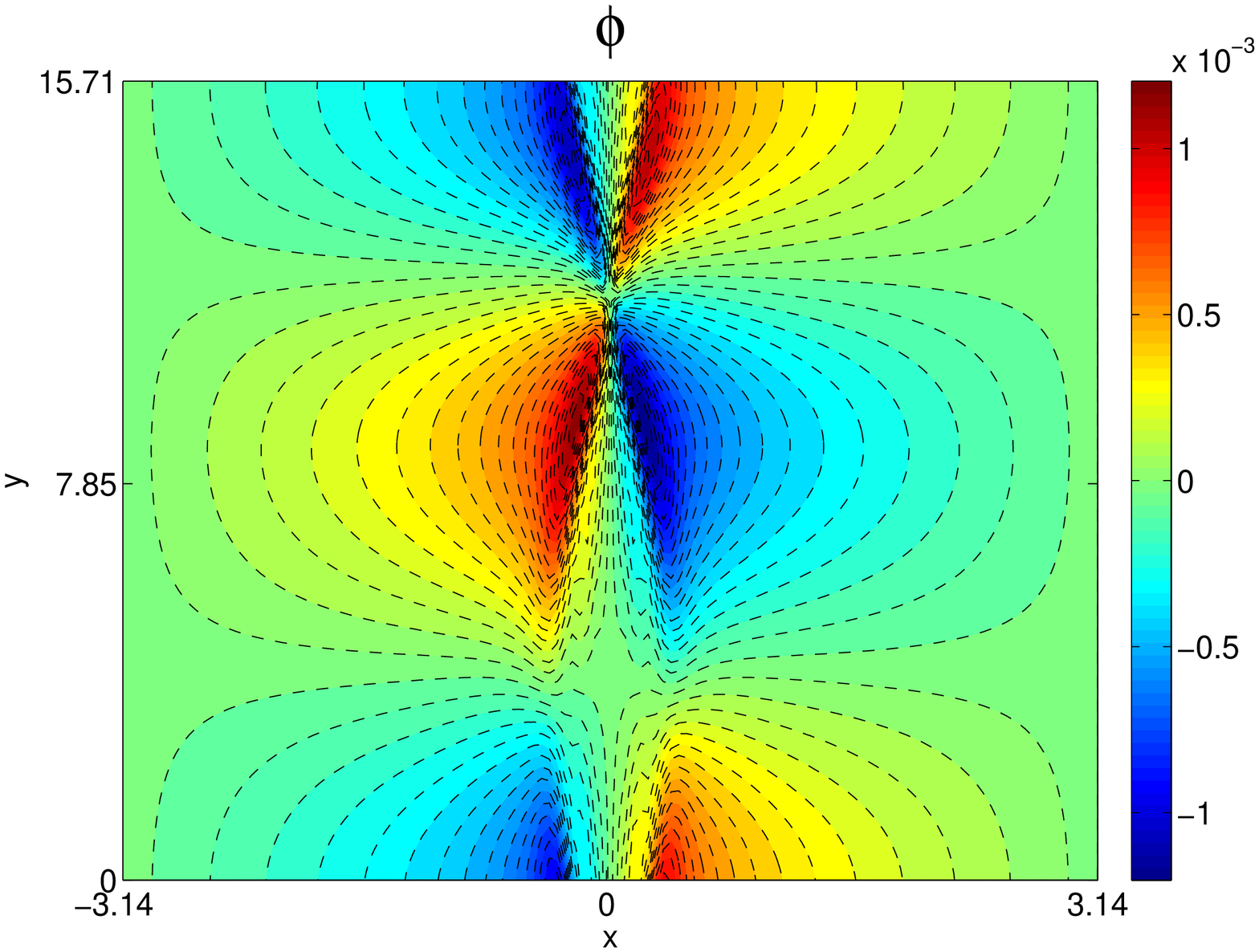}
\includegraphics[width=4.5cm,height=5cm]{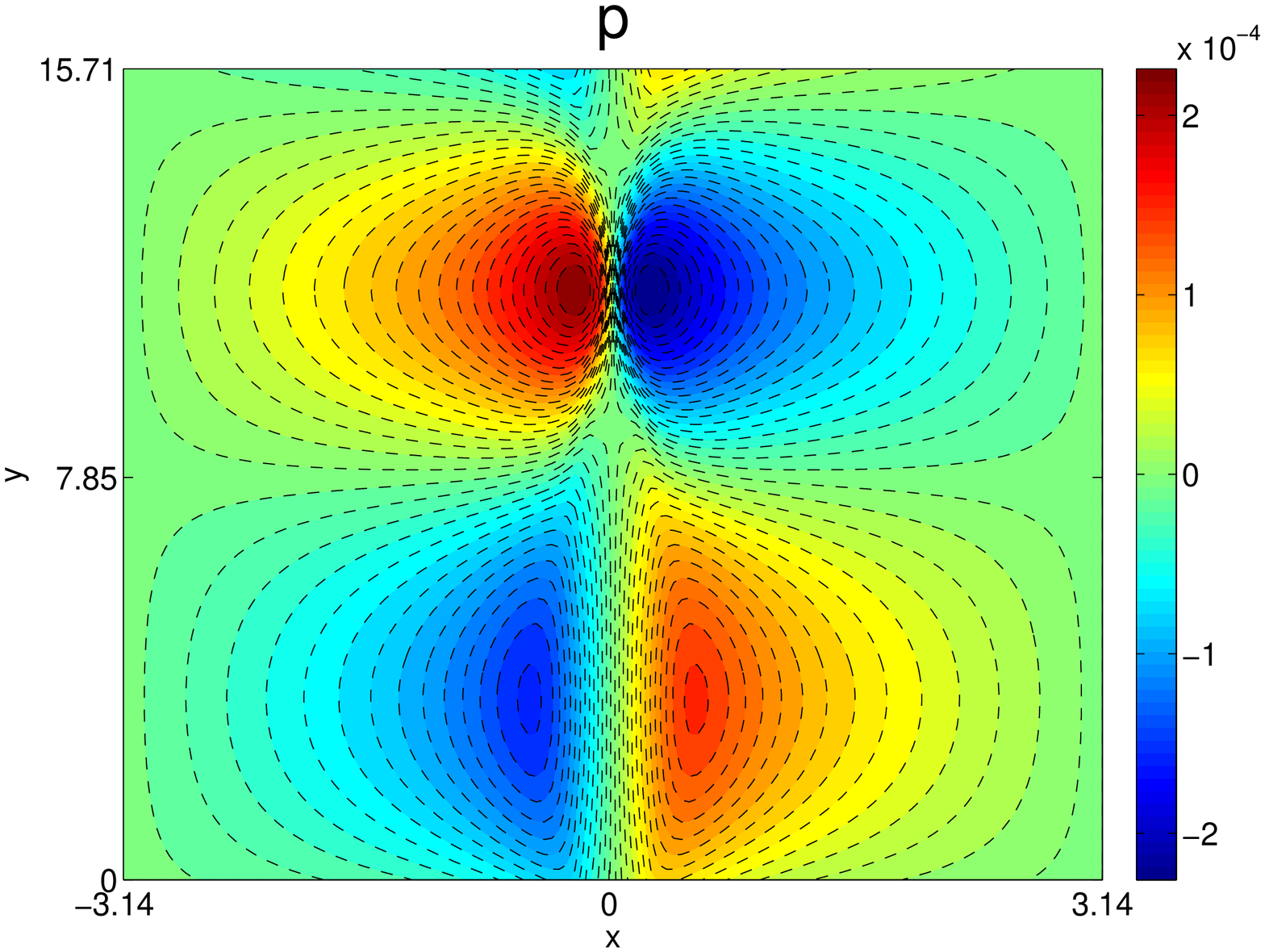}
\end{center}
\caption{Snapshots of the fields $\psi$, $\phi$ and $p$ before transition for $\beta=0.001$, at $t=5000\tau_A$.}
 \label{fig:figure4_b}
\end{figure}
\begin{figure}
\begin{center}
\includegraphics[width=4.5cm,height=5cm]{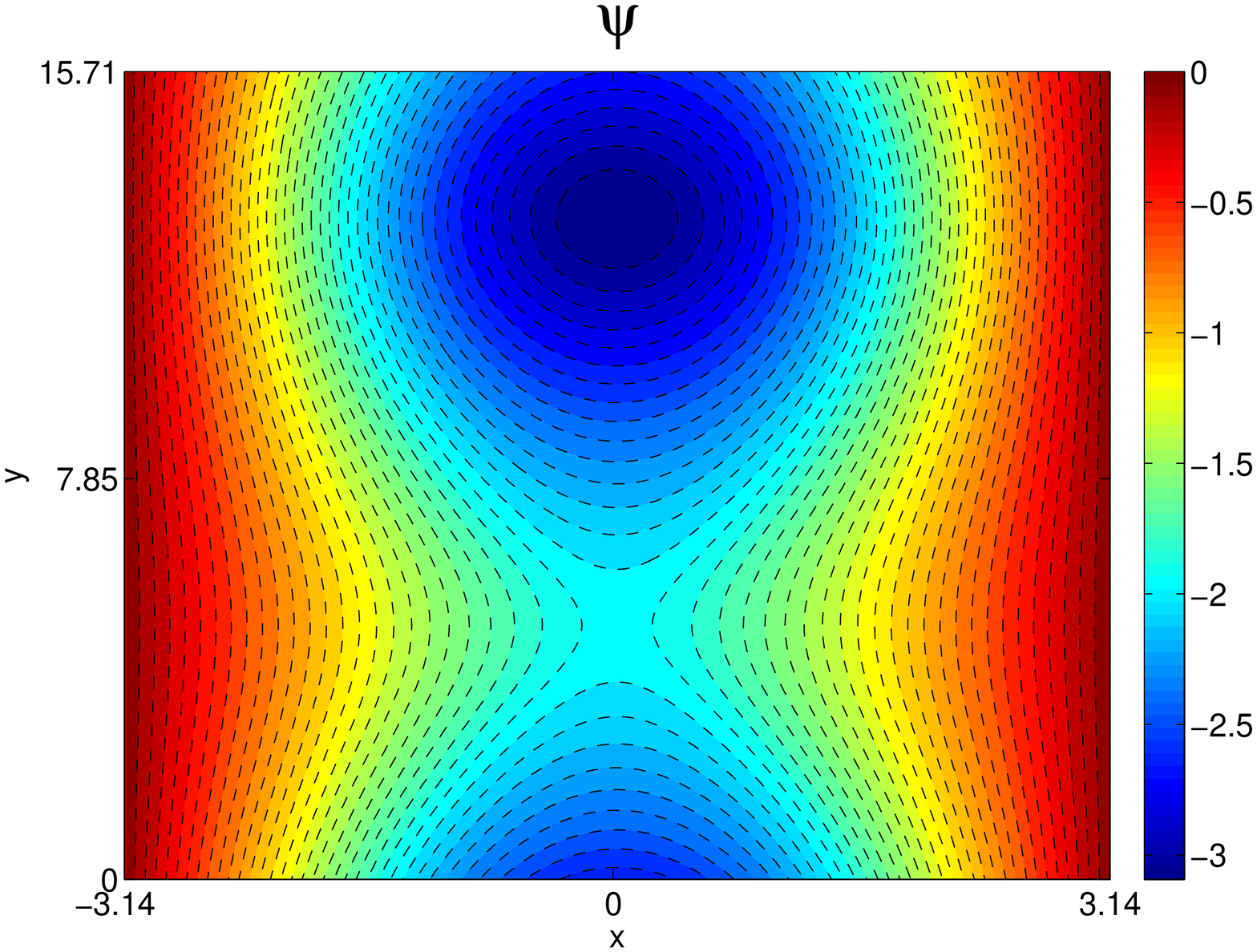}
\includegraphics[width=4.5cm,height=5cm]{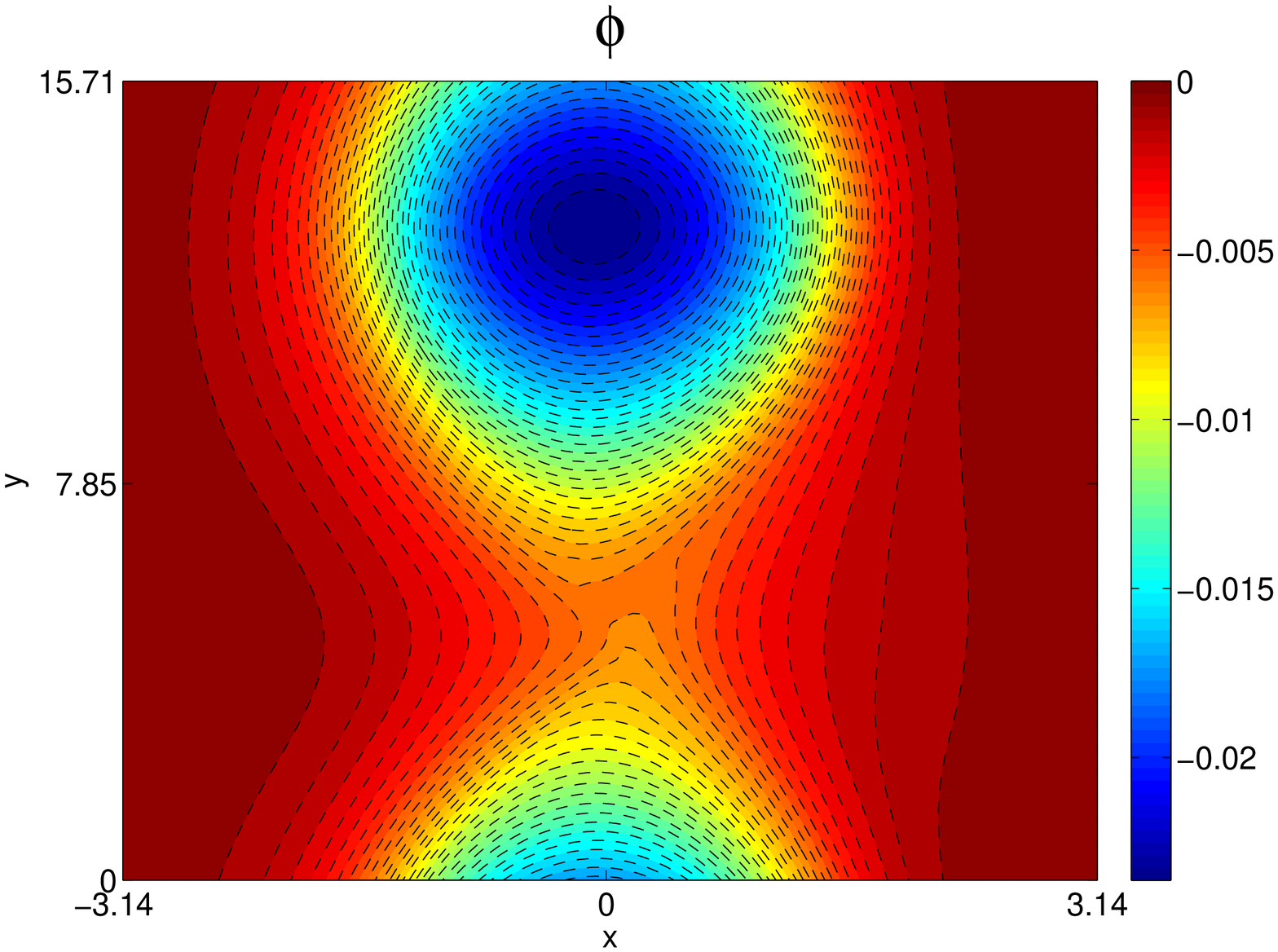}
\includegraphics[width=4.5cm,height=5cm]{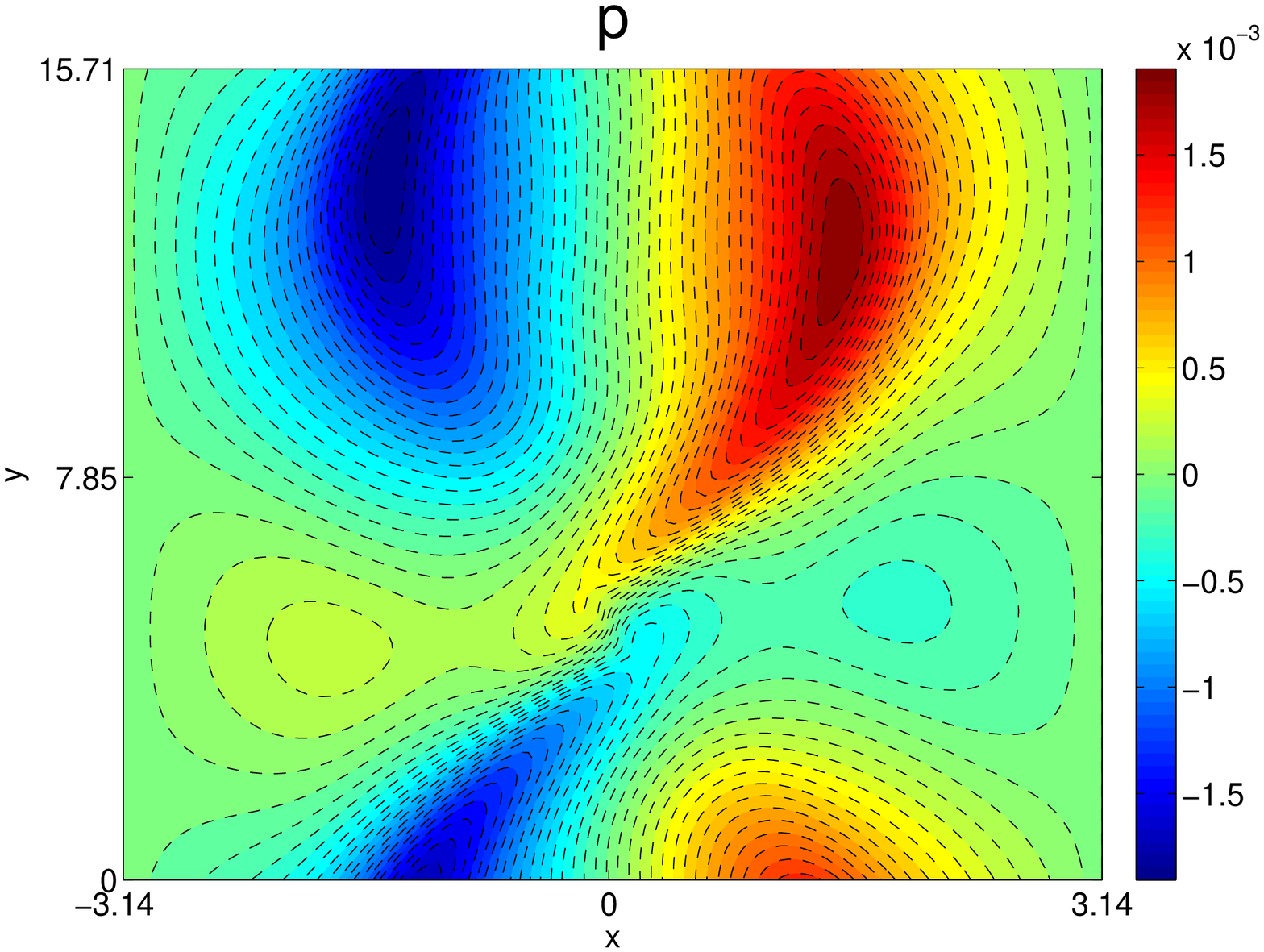}
\end{center}
\caption{Snapshots of the fields $\psi$, $\phi$ and $p$ after transition for $\beta=0.001$, at $t=18000\tau_A$.}
\label{fig:figure4_c}
\end{figure}
During the first two phases, the evolution of the energies is not strongly affected by the presence of the curvature terms. The evolution of the magnetic island follows closely the time trajectory of an island driven by a tearing instability. However, at $t_\star = 13200\;\tau_A$, a transition occurs.
Figures (\ref{fig:figure4_b}) and (\ref{fig:figure4_c}) show snapshots of the fields $\psi$, $\phi$ and $p$ respectively
before and after the transition. The two dimensional profiles of the pressure and the electrostatic potential (represented through isocontours) are strongly
affected by this transition. After this phase, the structure of the mode changes and a flattening of the pressure is obtained.


\subsection{Origin of the transition}

\begin{figure}
\begin{center}
\includegraphics[width=7cm,height=7cm]{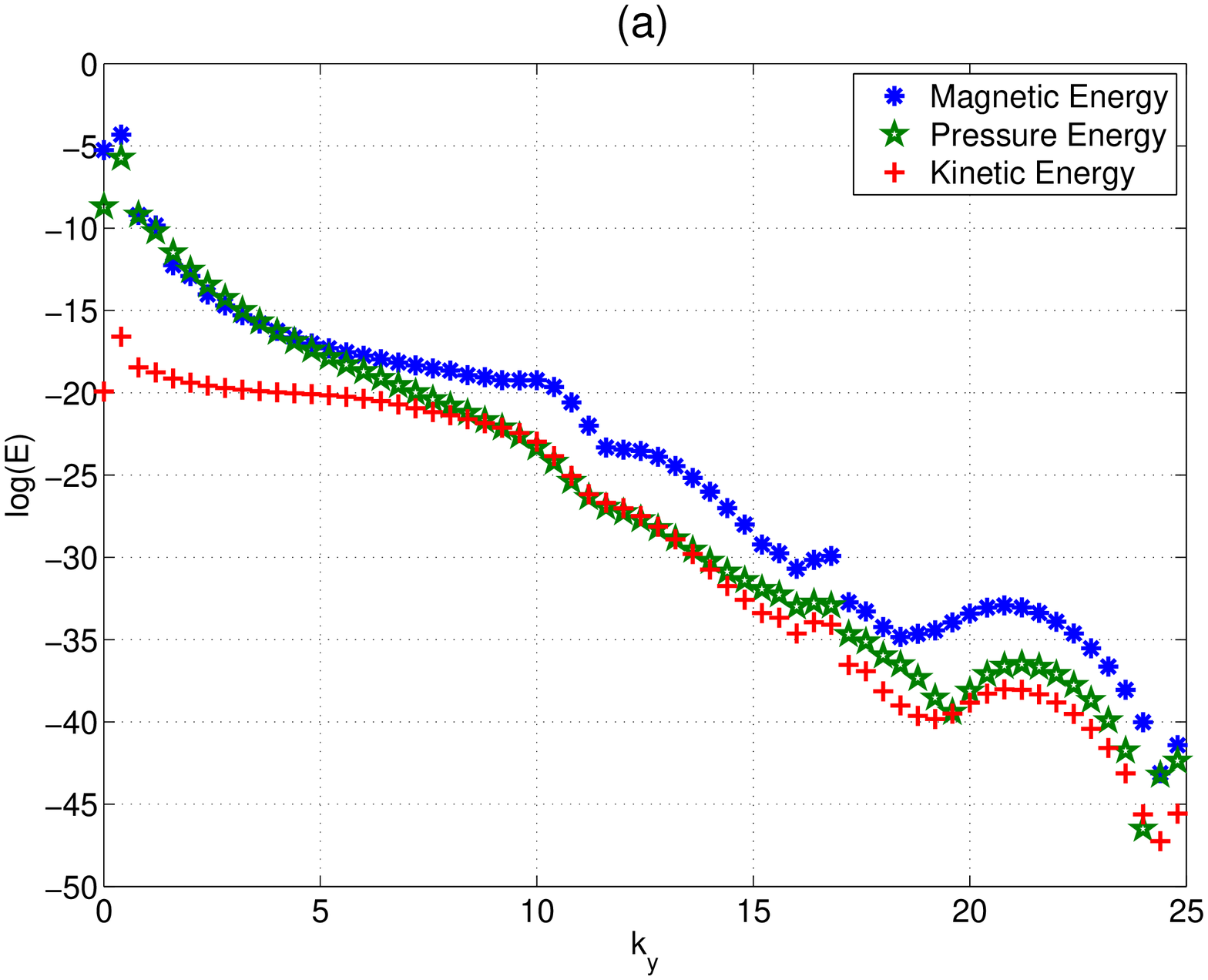}
\includegraphics[width=7cm,height=7cm]{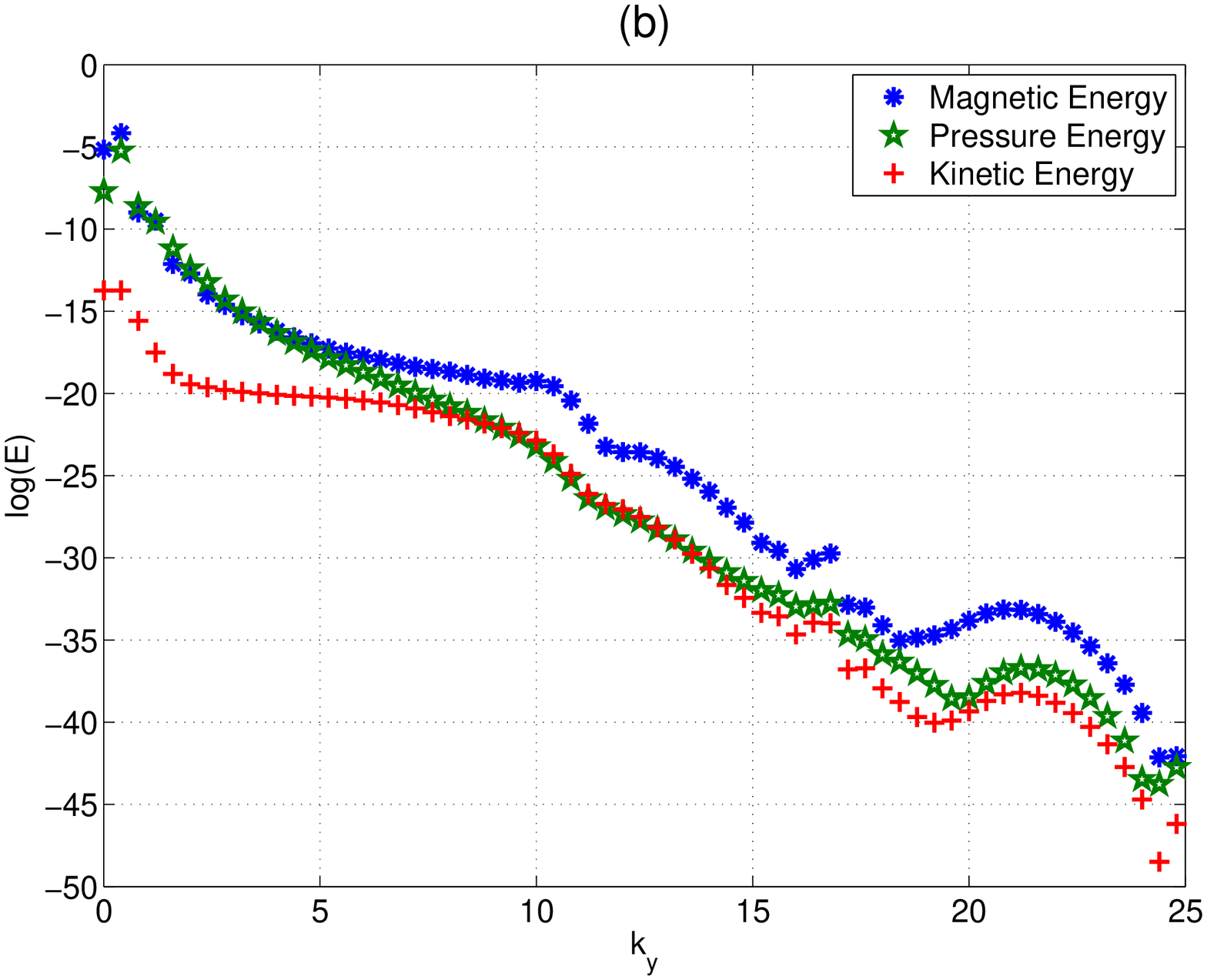}
\end{center}
\caption{Spectra before (a) and during the transition (b), at respectively $t=12000\tau_A$ and $t=14500\tau_A$.}
 \label{fig:figure4_d}
\end{figure}
\begin{figure}
\begin{center}

\includegraphics[width=7cm,height=7cm]{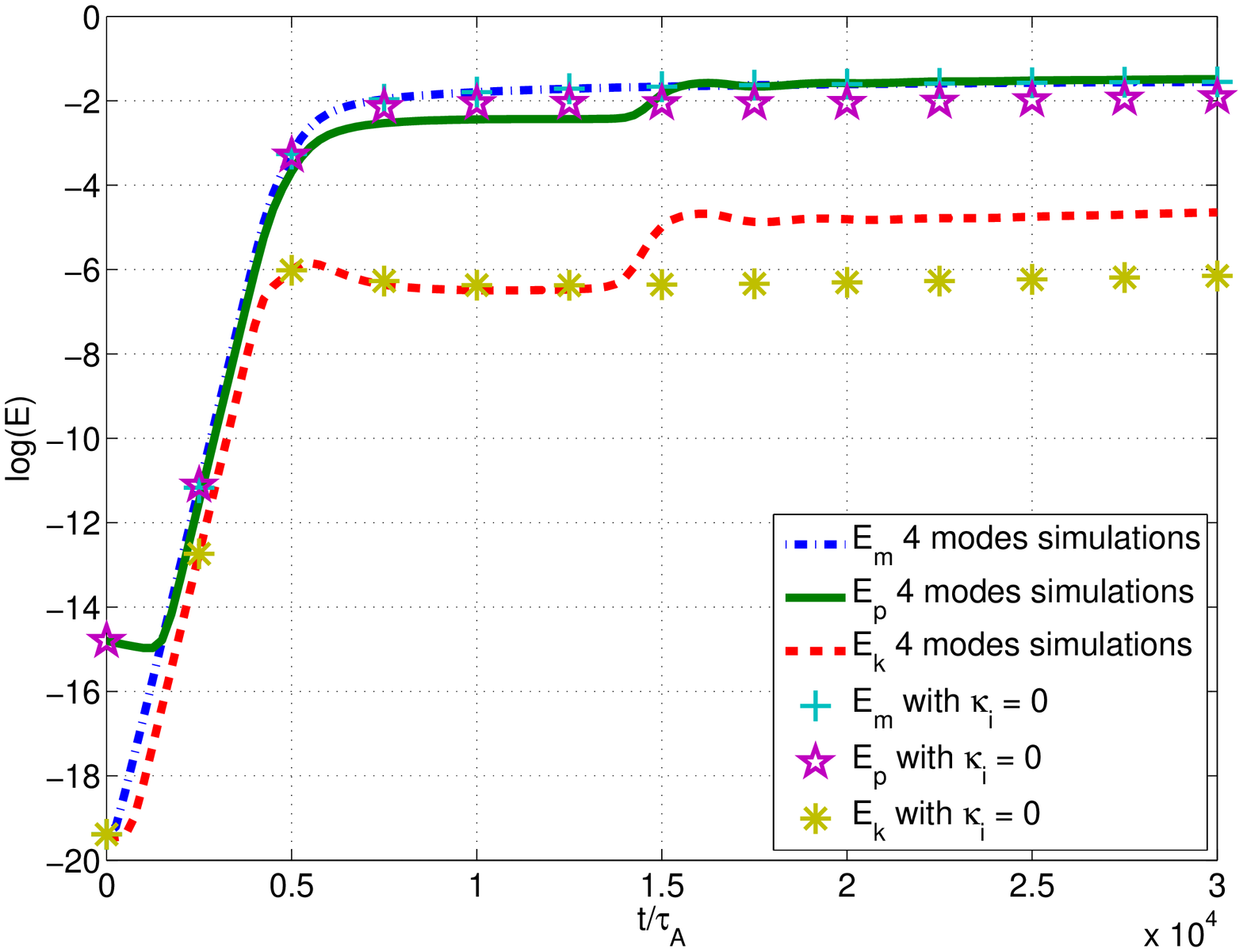}
\includegraphics[width=7cm,height=7cm]{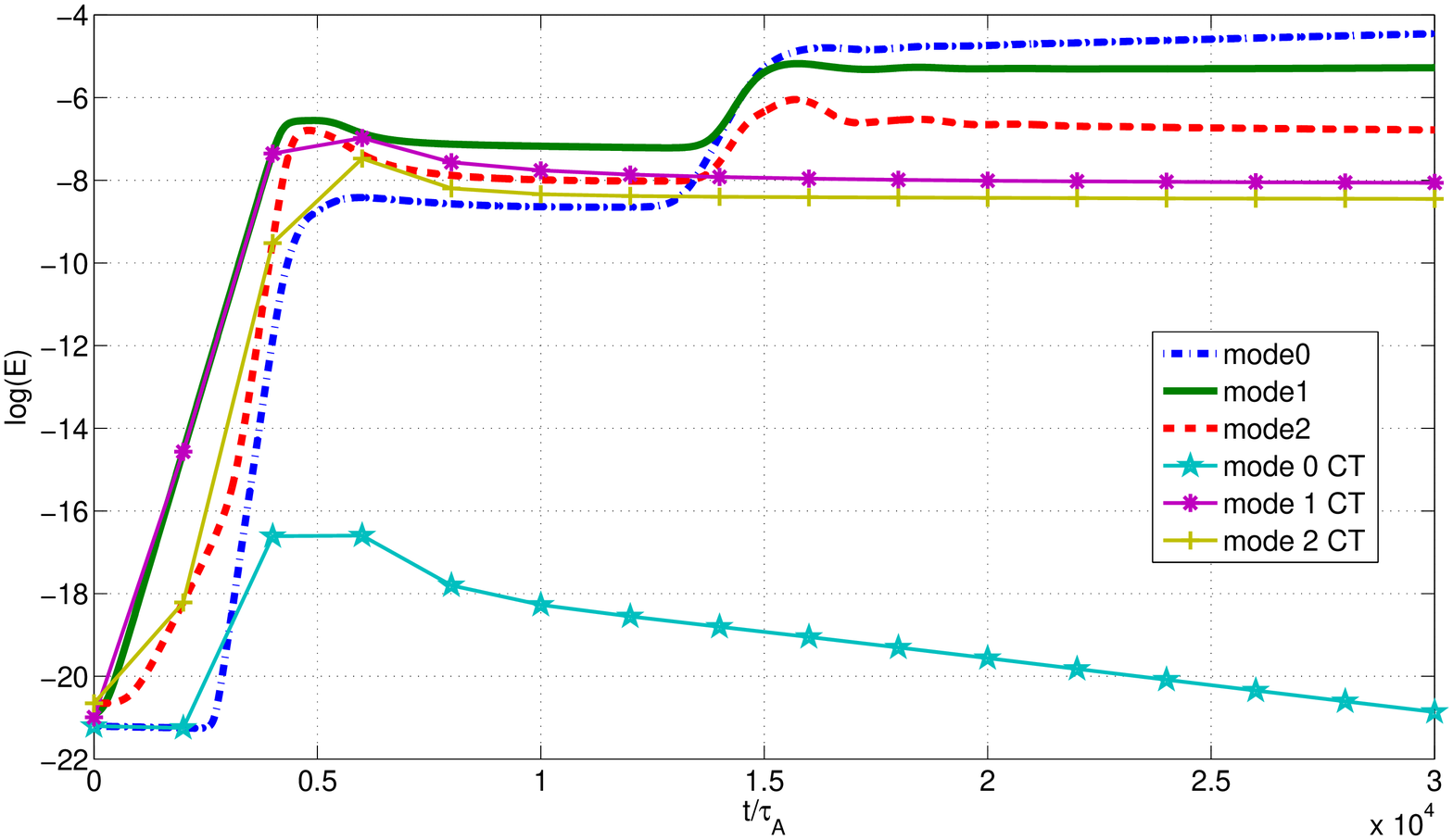}
\end{center}
\caption{ Runs with same parameters as figure (3) :  [Left] Time evolution of the energies with $\kappa_1=\kappa_2=0$ (4 modes runs). [Right]  Time evolution of the kinetic energy of modes 0, 1 and 2  along with the corresponding classical tearing case. }
 \label{fig:figure4_ef}
\end{figure}
In order to understand the origin of the transition and to characterize the structure of the electrostatic potential after the transition, we first assess the importance of small scales. The spectra before the transition (a) and during the transition (b) are shown in Figure (\ref{fig:figure4_d}). An equipartition between the energy of the magnetic flux and the energy of the pressure is observed at large scales $0.8<k_y<5$ whereas there is an equipartition between the energy of the pressure and the kinetic energy at small scales $8>k_y>14$. We observe that these properties continue to persist when the transition occurs, the energy level of the modes $k_y>1.2$ being roughly unaltered.
We also note that as apparent in Figure (\ref{fig:figure4_d}) only the large scales are affected at the transition. A detailed analysis shows that the transition occurs when the kinetic energy of the mode $k_y=0$ becomes equal to the one of $k_y=0.4$ (mode 1). This suggests that the dynamics of the structure of the electrostatic potential is quasi-linear and that the transition occurs when the mode $k_y=0$ becomes energetically dominant.
To delineate the quasi-linear nature of the magnetic island dynamics, we have performed a simulation with only four poloidal modes and with the same parameters ($\beta=10^{-3}$). Figure (\ref{fig:figure4_ef}) shows the time evolution of the energies  for this simulation. The comparison with Figure (\ref{fig:figure4_a}) demontrates that one needs only four modes to describe the time evolution of the energies. Therefore the dynamics of the system is the result of quasi linear effects.
Further, the time evolutions of the kinetic energies of modes 0, 1 and 2 are presented in the right hand side panel of Figure (\ref{fig:figure4_ef}) and compared with the evolution of modes in a classical tearing run, {\it i.e} without pressure effects ($p=0$ and $\kappa_i=0$). In the latter case, the growth is driven by the mode $1$, the transition does not occur and the mode $0$ is not generated. Conversely, in the case where the pressure effects are included, the transition occurs and the mode $k_y=0$, i.e zonal flow, is strongly generated. It is also the first to be amplified exponentially, at the beginning of the transition. This suggests that the transition is linked to a strong amplification of the zonal flow. Nevertheless, when we perform  the same run, suppressing artificially the mode 0, {\it i.e} the zonal flow, we find that the transition still occurs, roughly at the same time, but with a weaker amplitude.
\begin{figure}
\begin{center}
\includegraphics[width=4.5cm,height=5cm]{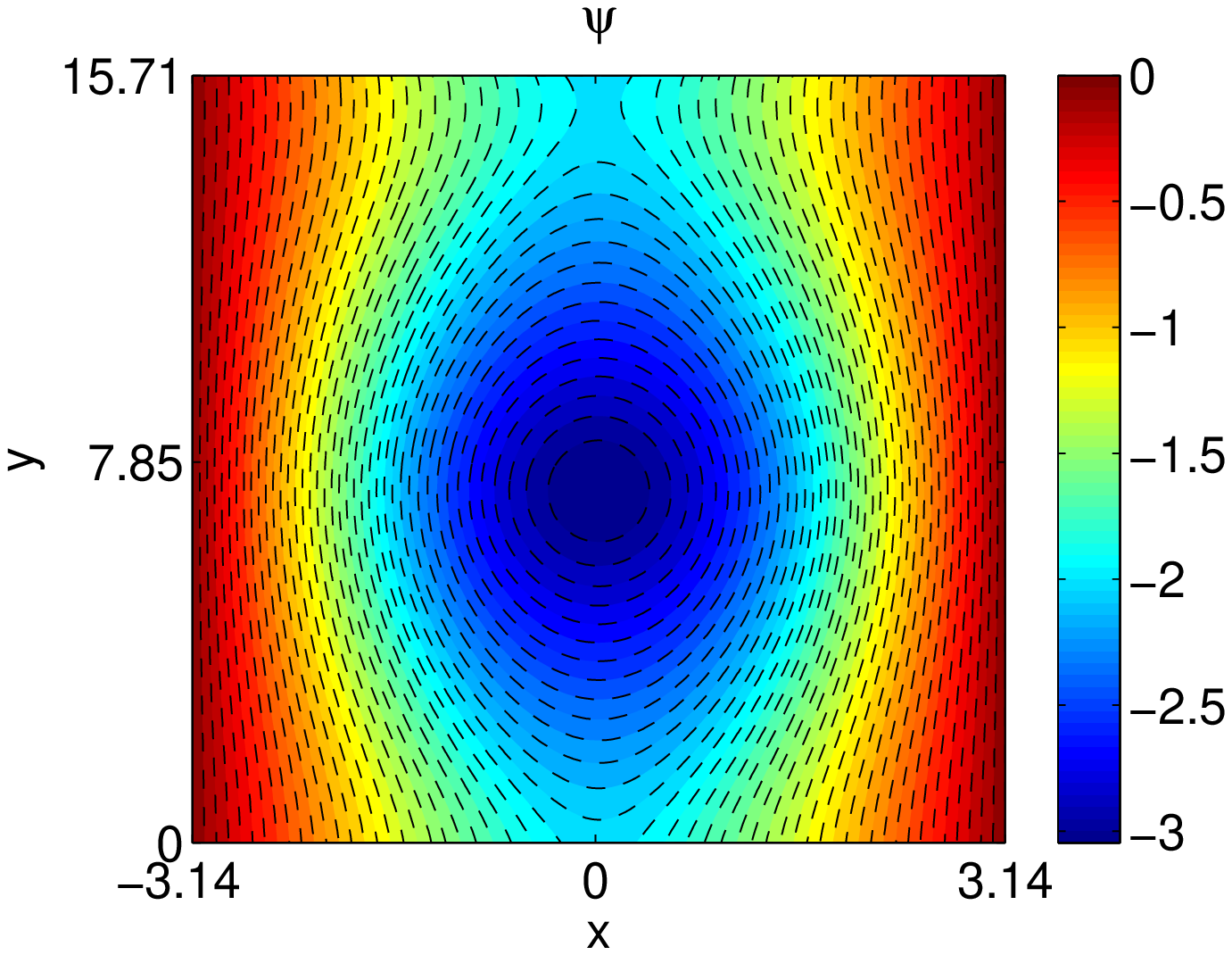}
\includegraphics[width=4.5cm,height=5cm]{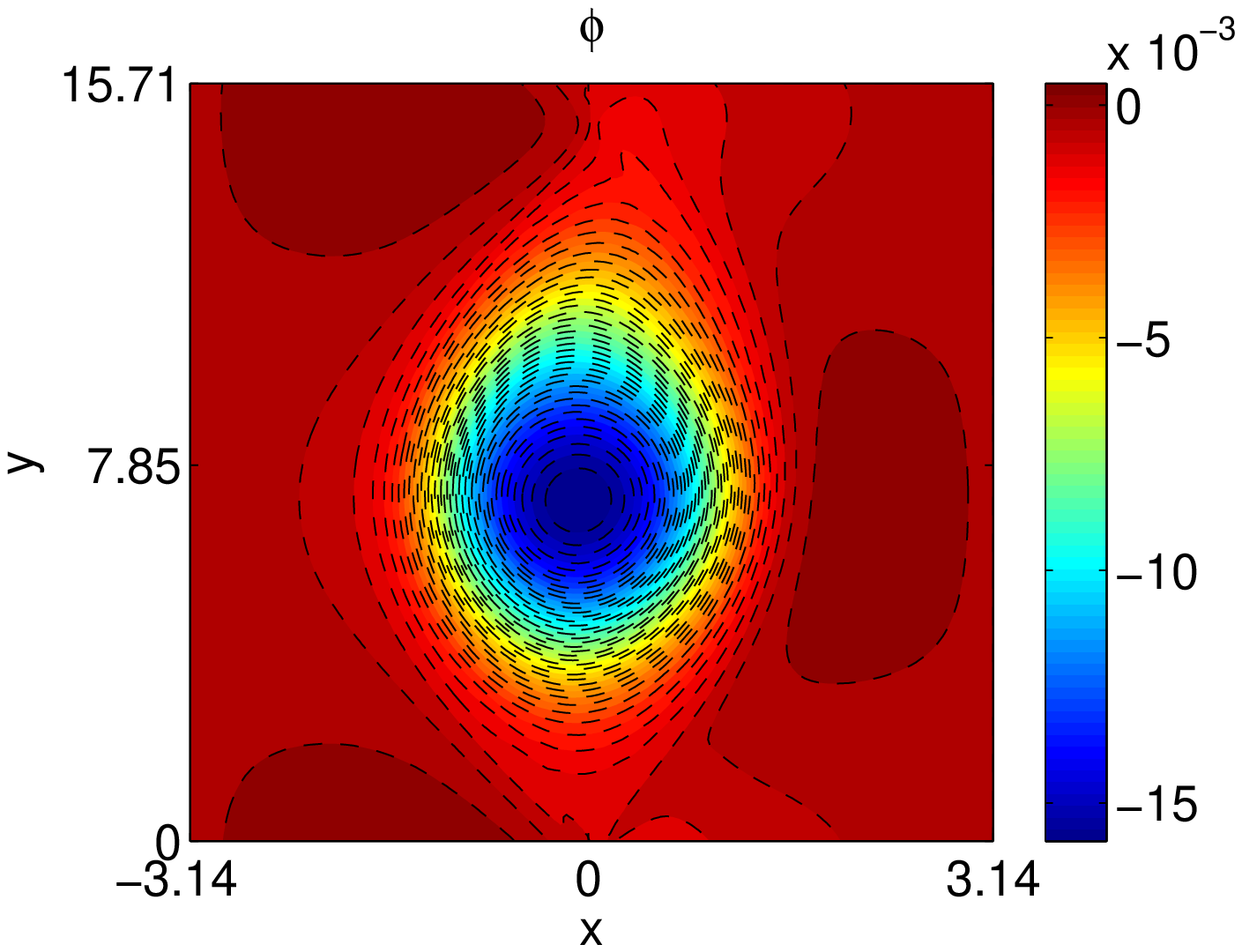}
\includegraphics[width=4.5cm,height=5cm]{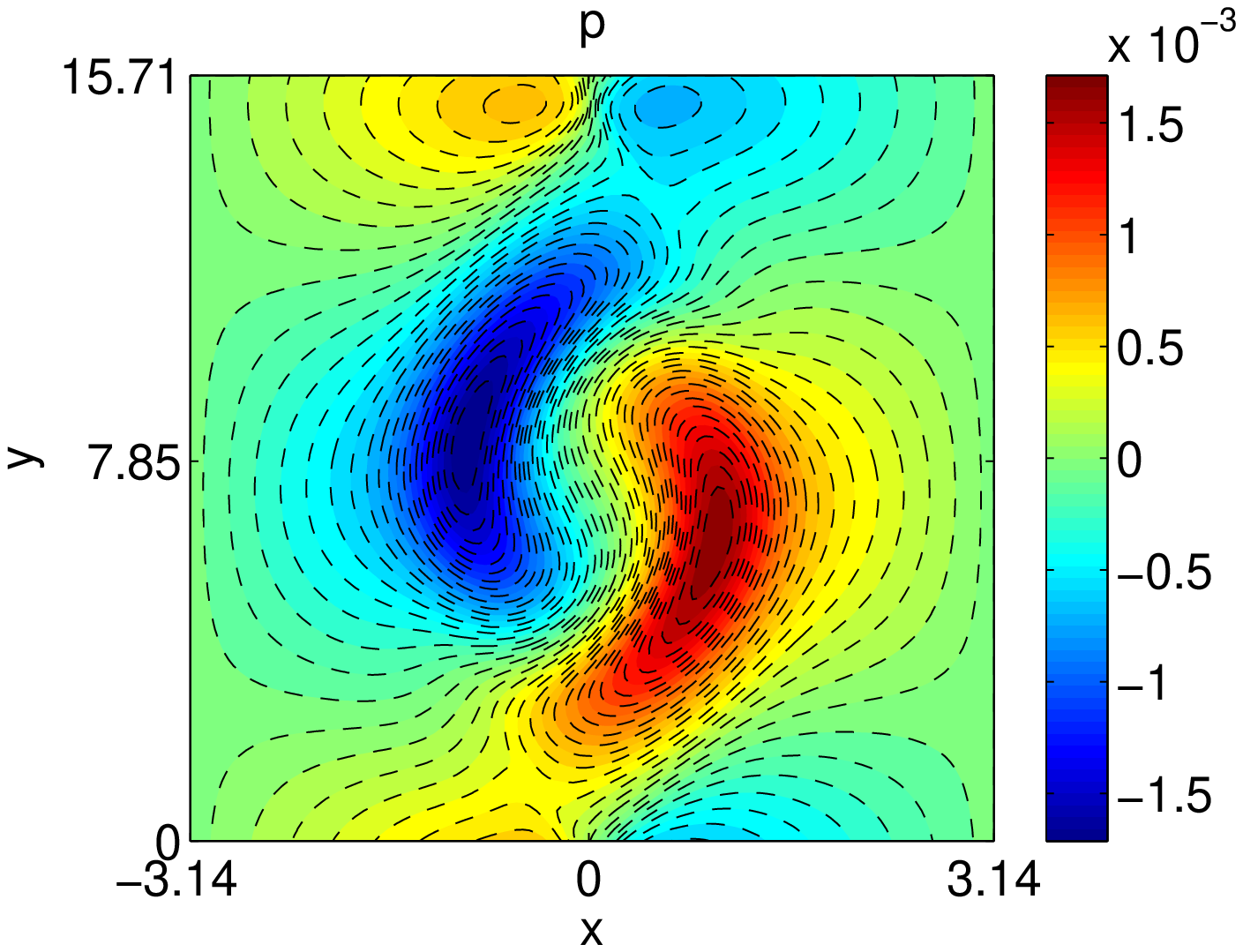}
\end{center}
\caption{Snapshots of the fields $\psi$, $\phi$ and $p$ for $\beta=0.001$, at $t=15000\tau_A$.}
\label{fig:figure4_efg}
\end{figure}
This suggests that althouh zonal flows play a predominant role they are not the sole factor responsible for the destabilizing mechanism. 
%
%
Indeed we can show from the analysis of the snapshots of $\phi$, $p$ and $\psi$ given by Figure (\ref{fig:figure4_efg}) that the mode 0 plays an important role in the triggering mechanism of the transition phase. We can observe that during the transition, the pressure cells are crossing the resonant surface at the current sheet in both directions, producing at the end a modification of the mode structure. Clearly the potential structure suggests that the crossing results from an advection by the flow. Let us stress that during this phase, the classical tearing picture of an incoming flow from the sheet into the island is no more valid. Between $t=15000\tau_A$ and $t=17000\tau_A$, $S(t)$ decreases because the reorganisation of the cell is radially equivalent to an exchange of pressure cells with the gradient of the pressure fluctuations being outward.
From an energetic point of view, see eq. (\ref{eq:energy}), $\kappa_1$ and $\kappa_2$ have a negligible effect because the dominant contribution in the interchange source term is linked to $\omega_\star/C^2$. In Figure (\ref{fig:figure4_ef}), the results of a simulation with four modes and without the curvature terms  are presented ($\kappa_1=\kappa_2=0$). We find that the transition does not occur.  At least, it does not occur at $2.5t_\star$, showing that a more complex mechanism due to curvature terms might be at play. We next investigate the origin of the zonal flow.

\subsection{Origin of the strong  increase of the zonal flow}

We investigate the origin of the zonal flow that is generated whenever $\beta\ne 0$ by considering separately  the energy transfer
from the tearing mode to the zonal flow and from also the other modes, in particular, the small scales. 

Following \cite{Ishizawa07}, the equation for the flow energy feeding the mode $k_m=m 2\pi/L_y$ can be written as,
\begin{equation}
\frac{d}{dt} E_{m}=T^R_m+T^M_m+T^{C}_m+T^{LB}_m+T^{KI}_m
\label{eq:EnergyTransfert}
\end{equation}
where
\begin{eqnarray*}
T^R_m &=& -\int dx \phi_m ([\phi,\omega])_m \mbox{ (Reynolds stress contribution)},\\
T^M_m &=&  \int dx \phi_m ([\psi,j])_m \mbox{ (Maxwell stress contribution)},\\
T^{C}_m &=&  -\kappa_1 k_m \int dx \phi_m p_m \mbox{ (curvature term contribution)},\\
T^{LB}_m &=& \int dx \phi_m ([\psi_0,j])_m \mbox{ (line bending term contribution)},\\
T^{KI}_m &=& \int dx \phi_m ([\psi,j_0])_m  \mbox{ (kink term contribution)}.\\
\end{eqnarray*}
Here $ \phi_m =   \hat\phi_m e^{i ky} +  \hat\phi_{-m} e^{-i ky} $, where $\hat\phi_m$ is the $m$ Fourier component of $\phi$.

\begin{figure}
\begin{center}
\includegraphics[width=7cm,height=5cm]{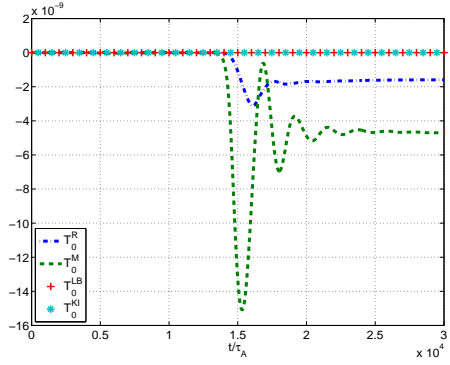}
\end{center}
 \caption{$\beta = 10^{-3}$: Time evolution of $T_0^R$, $T_0^M$, $T_0^{LB}$ and $T_0^{KI}$.}
\label{fig:figure4_hi}
\end{figure}
In order to understand the origin of the generation of the zonal flow, it is useful to project equation (\ref{eq:EnergyTransfert}) on the mode $m=0$:
\begin{equation}
\frac{d}{dt} E_0 = T^R_0+T^M_0+T^{LB}_0+T^{KI}_0
\label{eq:EnergyTransfert2}
\end{equation}
The curvature term does not directly feed  the zonal flow. In Figure (\ref{fig:figure4_hi}), the time evolutions of $T_0^R$, $T_0^M$, $T_0^{LB}$ and $T_0^{KI}$ for the simulation with $\beta = 10^{-3}$ are presented. The contributions of the line bending and the kink terms are very weak. However, at the transition, there is a strong generation of the Reynolds stress and the Maxwell stress contributions.
%
%
%
To proceed further, it is useful to separate the contributions from zonal flow ($m=0$), the mode $m=1$ and other modes for each of the transfer functions. Let us introduce
 $\phi_{>1}=\sum_{m>1}\phi_m$. We can define three contributions in each transfer function. For instance, in the case of $T^R_m$, we have:
\begin{equation}
T^R_m = T^R_{0m} + T^R_{1m} +  T^R_{>1m}
\end{equation}
where $T^R_{0m}= \int dx \phi_m ([\phi_0,\omega_0])_m$,  $T^R_{1m}= \int dx \phi_m ([\phi_1,\omega_1])_m$,  and $T^R_{>1m}= \int dx \phi_m ([\phi_{>1},\omega_{>1}])_m$. Clearly, by definition $T^R_{11}=0 $ and $T^R_{0m}=T^M_{0m} =0$.
Let us focus on the energy transfer to zonal flow, neglecting the weak contributions of the line bending and the kink terms. Equation (\ref{eq:EnergyTransfert2}) then becomes:
\begin{equation}
\frac{d}{dt} E_{0}=T^R_{10} +  T^R_{>10}+T^M_{10} + T^M_{>10}
\end{equation}
Using the above prescription we have checked that the main contibution to the Reynolds and Maxwell stresses comes from the mode $m=1$ while the contributions of the small scales are weak.

\subsection{Effect of the $\beta$ parameter on the nonlinear dynamics}
\begin{figure}
\begin{center}
\includegraphics[width=7cm,height=7cm]{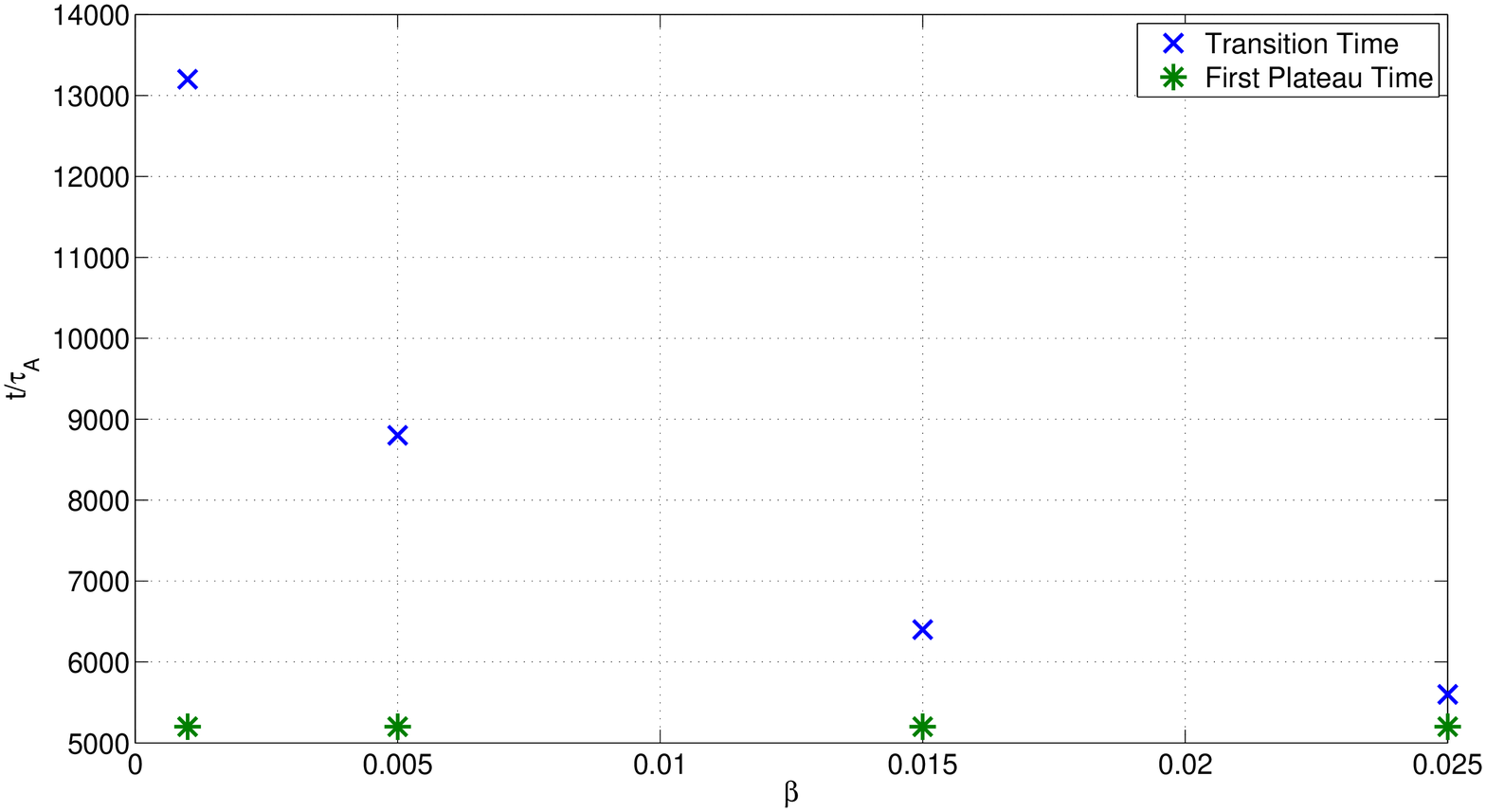}
\includegraphics[width=7cm,height=7cm]{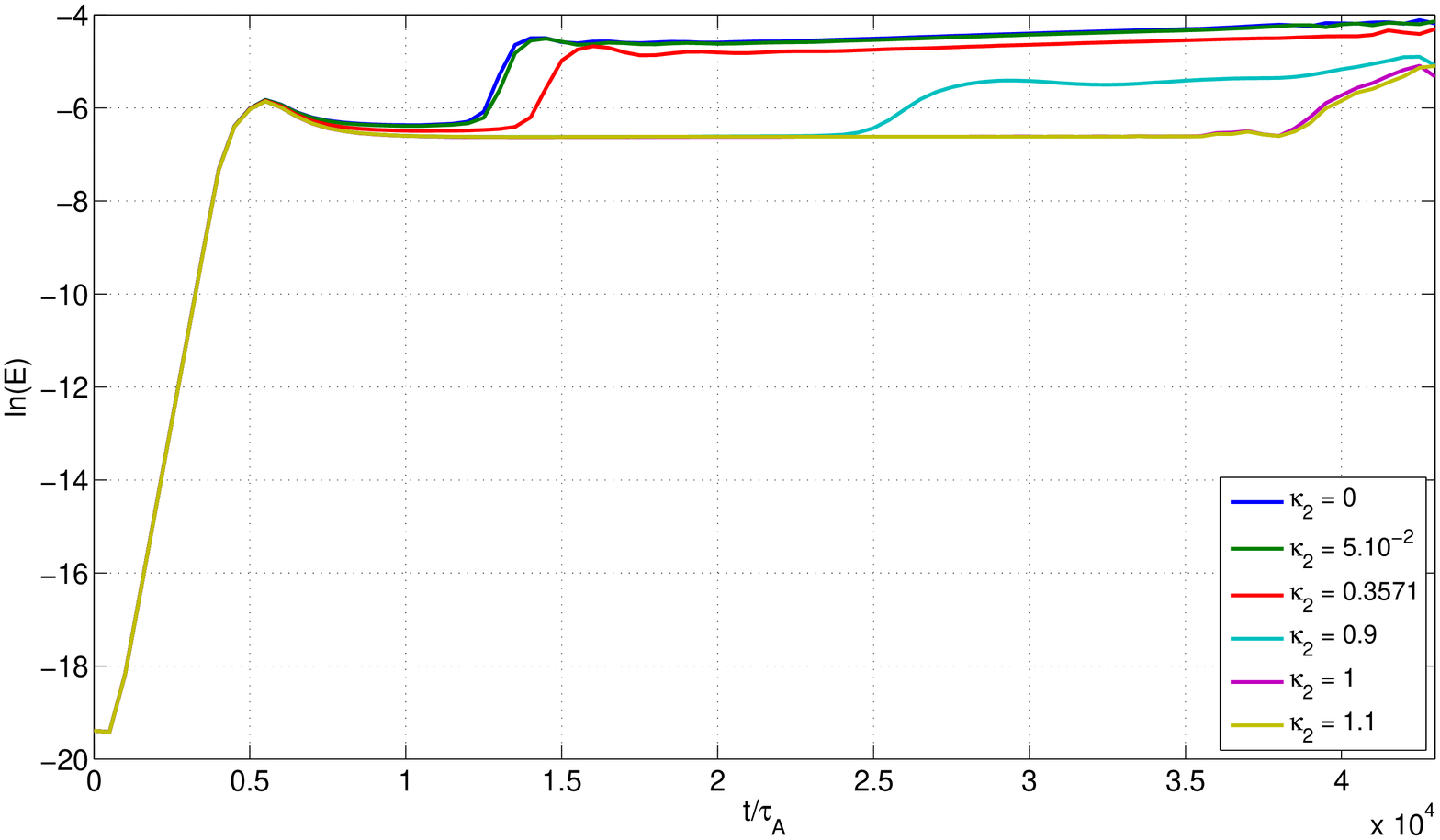}
\end{center}
\caption{Effect of the $\beta$ parameter and of $\kappa_2$ on the transition : [Left] Time at the beginning of the first quasi plateau phase and time at the beginning of the transion versus $\beta$. [Right] Time evolution of the kinetic energy for simulations with $\beta=10^{-3}$ but with different values of $\kappa_2$.}
 \label{fig:figure4_gj}
\end{figure}

The left panel of Figure (\ref{fig:figure4_gj}) presents the transition time and the time where the first quasi plateau saturation occurs for various values of $\beta$. 
%
We note that the time corresponding to the first quasi-plateau phase does not depend on the $\beta$ parameter. This is in agreement with the results of Figure (\ref{fig:figure4_a}) which shows the time evolution of the energies for two cases namely the classical tearing case ($\beta=0$) and the $\beta=10^{-3}$ case. However the saturation time depends strongly on the $\beta$ parameter. For a regime where the pressure effects are strong, i.e for a high value of $\beta$, the transition occurs quickly whereas, for low $\beta$ regimes, the transition occurs later. 
Further, in the right panel of Figure (\ref{fig:figure4_gj}), the effect of the interchange parameter $\kappa_2$  on the transition is shown for a simulation with four modes and with $\beta = 10^{-3}$. The transtion time depends clearly on the $\kappa_2$ parameter. It tends to stabilize the first plateau phase. 
%

%
To summarize, the nonlinear transition results from quasilinear effects. Zonal flow amplification at the transition is due to the energy transfer from the mode $m=1$ to the mode $m=0$ through mainly the Maxwell stress. The curvature term $\kappa_2$ linked to the interchange effect does not directly feed the growth of the zonal flow. However, we have shown that this term controls the transition time $t_\star$. The transition leads to an effective radial exchange of pressure cells generating an outward mean pressure gradient of fluctuations. The shape of the pressure structure after the transition implies that a diamagnetic velocity $\tilde\omega_\star$ has been nonlinearly generated, driving a rotation of the island. This driving is of course in competition with the zonal flow. Let us in the next section analyze quantitatively  the island poloidal rotation.


\section{Study of the island poloidal rotation}
\subsection{Model for the island rotation frequency}

Following \cite{Nishimura08} where a study of the rotation frequency of the island has been done for the case of a drift-tearing mode, we investigate the origin of the magnetic island poloidal rotation. Let us project the Ohm's law (Eq.\ref{eq:equPSI}) on the mode $m=1$ assuming that for the mode $k_y$, $\tilde{\psi}_{k_y} (x,y,t)= \psi_{k_y}(x)e^{ik_yy}e^{-i\gamma t}$, where the real part of $\gamma$ is the frequency of the island rotation and the imaginary part is the linear growth rate of the island. Neglecting the nonlinear contribution of the modes $k_y>k_1$, we obtain the expression for the rotation frequency $\omega$ of the island
\begin{equation}
\omega = \omega^\star+\tilde{\omega}^\star+\tilde{\omega}_{E\times B}+L_{\psi_0}+L_\eta\label{eq:equFREQUENCY}
\end{equation}
where
\begin{eqnarray*}
\omega^\star &=& k_1v^\star  ,\\
\tilde{\omega}^\star &=& -k_1\partial_x p_0 ,\\
\tilde{\omega}_{E\times B} &=& k_1\partial_x \phi_0 ,\\
L_{\psi_0} &=& -Re\left(k_1\psi'_0\frac{\phi_{k_1}\left(x\right)-p_{k_1}\left(x\right)}{\psi_{k_1}\left(x\right)}\right) ,\\ 
L_\eta &=& Re\left(i\eta\frac{\left(\partial^2_x-k^2_1\right)\psi_{k_1}\left(x\right)}{\psi_{k_1}\left(x\right)}\right),\\ 
\end{eqnarray*}
\noindent
are respectivly the linear diamagnetic drift, the nonlinear diamagnetic drift, the contribution of the equilibrium magnetic field and the contribution of the resistivity.
In general, each term of Eq.(\ref{eq:equFREQUENCY}) is not a constant inside the current sheet $\delta$, so we consider their radial averages over the current sheet to contribute to the rotation frequency. Eq. (\ref{eq:equFREQUENCY}) becomes :
\begin{equation}
<\omega>_\delta = < \omega^\star+\tilde{\omega}^\star>_\delta+<\tilde{\omega}_{E\times B}>_\delta+<L_{\psi_0}>_\delta+<L_\eta>_\delta\label{eq:equFREQUENCY2}
\end{equation}
where $<.>_\delta$ means an average over the current sheet.
\begin{figure}
\begin{center}
\includegraphics[width=10cm,height=7cm]{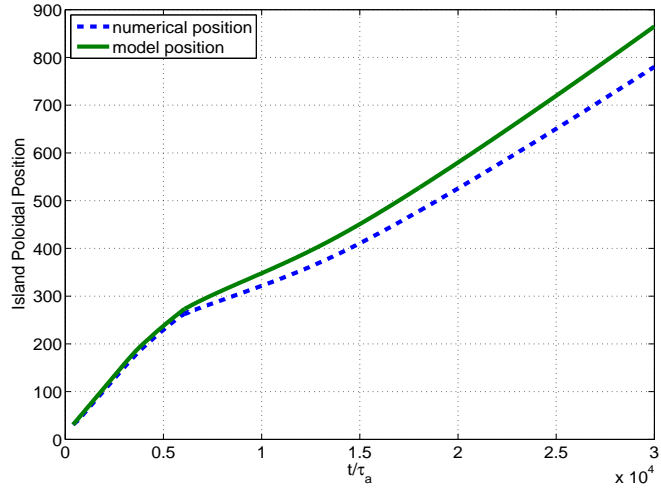}
\end{center}
\caption{Time evolution of the poloidal position of the island : comparison between the model and numerical data for $\beta=0.025$.}
 \label{fig:figure5_a}
\end{figure}

\begin{figure}
\begin{center}
\includegraphics[width=7cm,height=7cm]{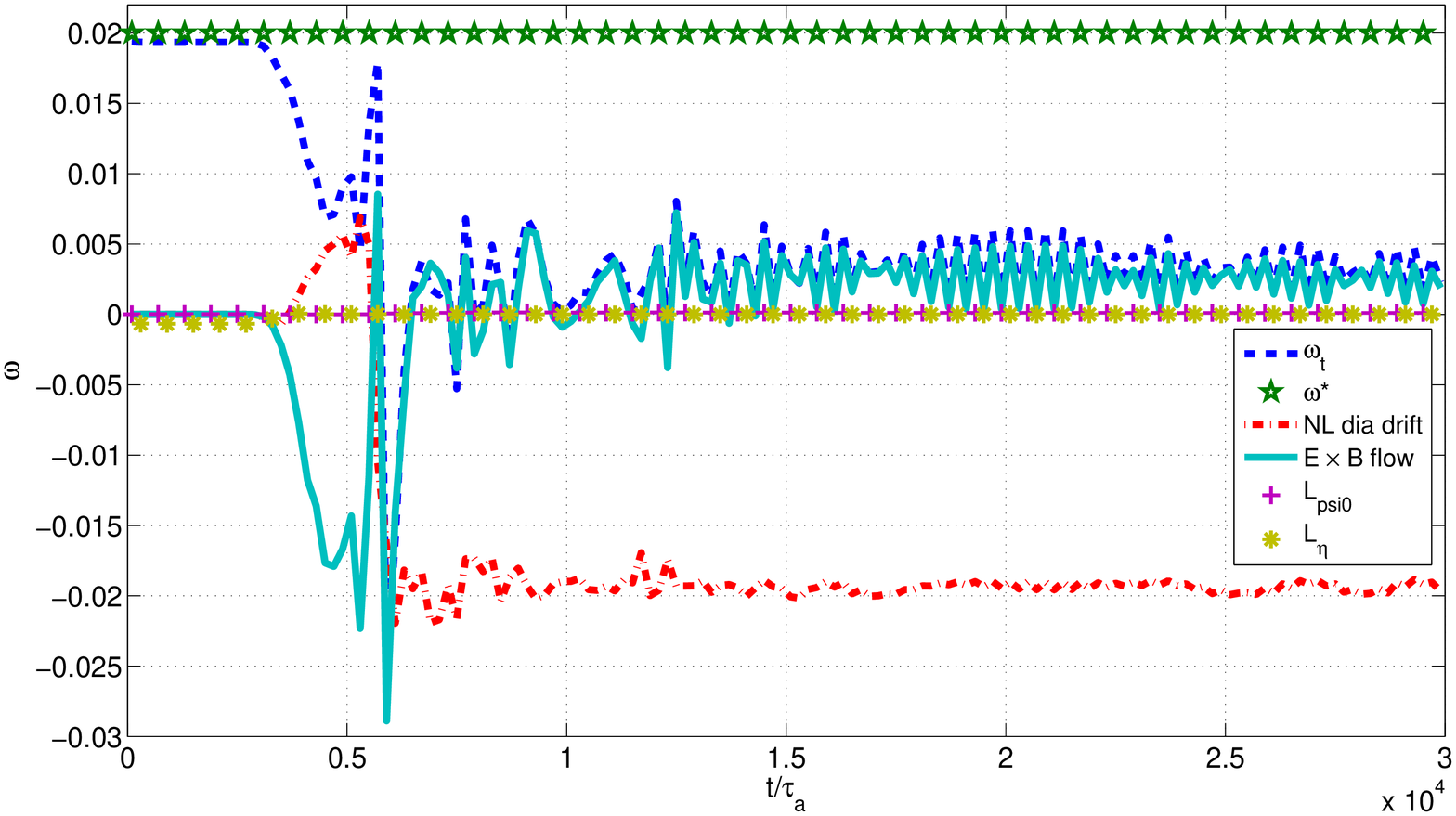}
\includegraphics[width=7cm,height=7cm]{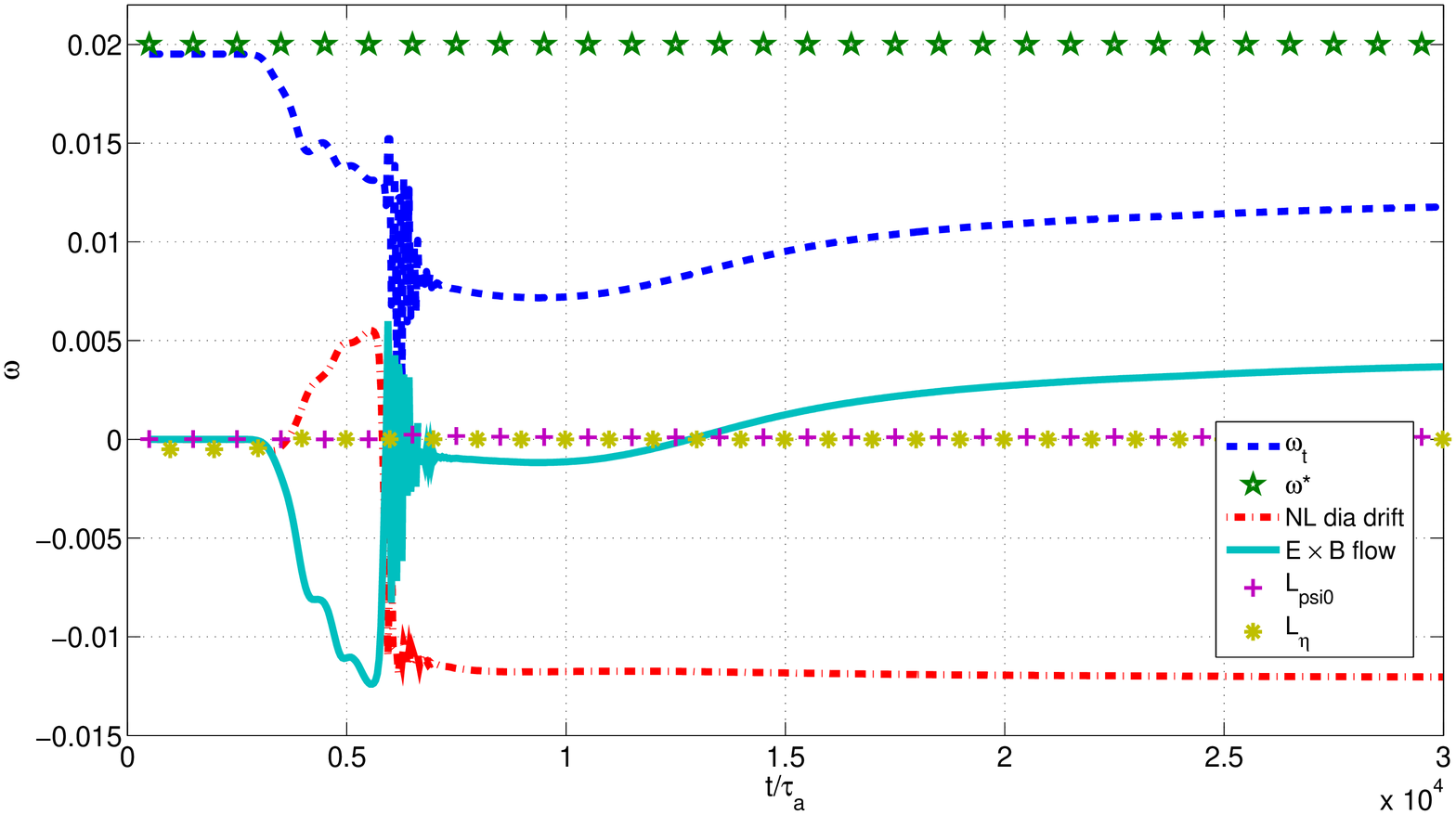}
\end{center}
\caption{Time evolution of the frequencies for simulations with $\beta=0.025$ : [Left] Simulation setting $\kappa_{1}=\kappa_{2}=0$. [Right] Simulation including the interchange terms.}
 \label{fig:figure5_b}
\end{figure}

In Figure (\ref{fig:figure5_a}), the time evolution of the poloidal position is presented for a nonlinear simulation with $\beta = 0.025$. The dynamics of the energies for a simulation with $\beta=0.025$ has the same behaviour as the one obtained in Figure (\ref{fig:figure4_a}) with $\beta=10^{-3}$. However, as  shown in the left panel of Figure (\ref{fig:figure4_gj}), with such a high value of $\beta$, the transition occurs earlier around $t=5600\tau_A$. In Figure (\ref{fig:figure5_a}) a comparison with the island position obtained from the model Eq.(\ref{eq:equFREQUENCY2}) is also shown. The derived model is in agreement with the numerical data and the dynamics of the island rotation is recovered. We would like to mention here that the time integration for these results has been performed on a very long time scale compared with the Alfv\'en time.\\

Equation (\ref{eq:equFREQUENCY2}) shows that the effective island frequency is the result of different contributions. In order to investigate the effect of the interchange terms on the island rotation, Figure (\ref{fig:figure5_b}) presents the evolution of each frequency for a simulation with $\beta=0.025$ setting $\kappa_1=\kappa_2=0$ (left panel) and for a simulation with $\beta = 0.025$ including the curvature/interchange terms (right panel). First, for the two simulations, the contributions to the rotation of the equilibrium magnetic field $L_{\psi_0}$ and of the resistivity $L_\eta$ are weak. Moreover, Figure (\ref{fig:figure5_b}) shows that the frequency dynamics is not affected by the curvature terms during the linear regime and during the first quasi plateau phase. Actually, during the linear formation of the magnetic island, the rotation is controlled mainly by the linear diamagnetic drift while $\tilde{\omega}^\star$ and $\tilde{\omega}_{E\times B}$ are weak. During the first quasi plateau phase, the nonlinear diamagnetic drift and the $E\times B$ poloidal flow are strongly generated, and affect the island rotation. The $E\times B$ poloidal flow is the most important contribution to the frequency during this regime. However, after the first quasi plateau phase, interchange terms affect the dynamics of the frequency. When the interchange terms are switched off, the time evolution of the frequencies is in agreement with previous results found for a drift tearing mode \cite{Nishimura08}. After the nonlinear generation of the flows, linear and nonlinear diamagnetic drifts cancel each other. As a result, the $E\times B$ poloidal flow controls the effective frequency of the island,  $\omega_t\sim\tilde\omega_{E\times B}$. The right panel of Figure (\ref{fig:figure5_b}) shows that when interchange terms are included, such canceling of the total diamagnetic frequency does not occur anymore after the transition. Hence the total diamagnetic drift then provides the main contribution to the island rotation. However, after the transition, clearly 
$\partial \omega_{t}/\partial t \sim \partial \omega_{E\times B}/\partial t $.
The asymptotic island velocity is enhanced by the curvature terms $\kappa_1$ and $\kappa_2$.  

\subsection{Origin of the $E\times B$ flow}

Nonlinearly, the $E\times B$ flow is generated and affects the rotation of the island.
In order to investigate its origin, the flow equation (Eq.\ref{eq:equPHI}) is projected on the mode $k_1$ for the limiting case of $\nabla^2_\perp\approx\partial^2_x$. We obtain :
\begin{equation}
\partial_t\partial_x^2\phi_0 = -\frac{1}{L_y}\int_{L_y}[\phi,\nabla_\perp^2\phi]dy+\frac{1}{L_y}\int_{L_y}[\psi,\nabla_\perp^2\psi]dy+\frac{\mu}{L_y}\int_{L_y}\partial_x^4\phi dy\label{eq:flow1}
\end{equation}

We have defined the $E \times B$ poloidal flow as $\tilde{\omega}_{E\times B}= k_1\partial_x \phi_0$. So multiplying Eq. (\ref{eq:flow1}) by $k_1$ and averaging over the current sheet $\delta$, we obtain :
\begin{equation}
\partial_t<\tilde{\omega}_{E\times B}>_\delta = R(t)+M(t)+V(t)\label{eq:flow2}
\end{equation}
where
\begin{eqnarray*}
R\left(t\right) &=& -\frac{k_1}{\delta L_y}\int_\delta\int_{L_y}[\phi,\nabla_\perp^2\phi]dydx,\\
M\left(t\right) &=&   \frac{k_1}{\delta L_y}\int_\delta\int_{L_y}[\psi,\nabla_\perp^2\psi]dydx,\\
V\left(t\right) &=& \frac{\mu k_1}{\delta L_y}\int_\delta\int_{L_y}\partial_x^4\phi dydx
\end{eqnarray*}
with $R\left(t\right)$ being the Reynolds stress, $M\left(t\right)$ being the Maxwell stress and $V\left(t\right)$ being the viscosity contribution to the $E\times B$ flow.\\

\begin{figure}
\begin{center}
\includegraphics[width=7cm,height=7cm]{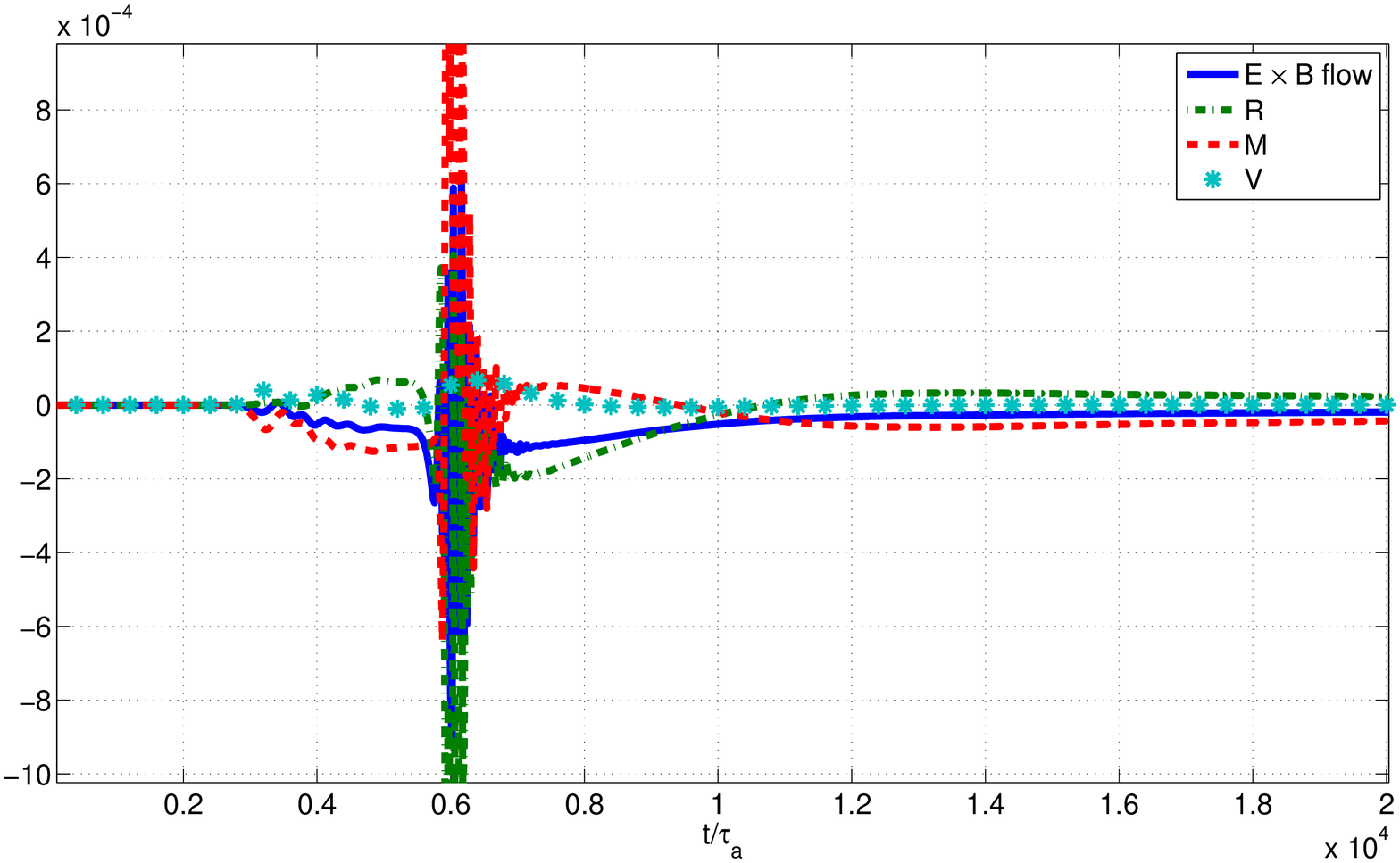}
\includegraphics[width=7cm,height=7cm]{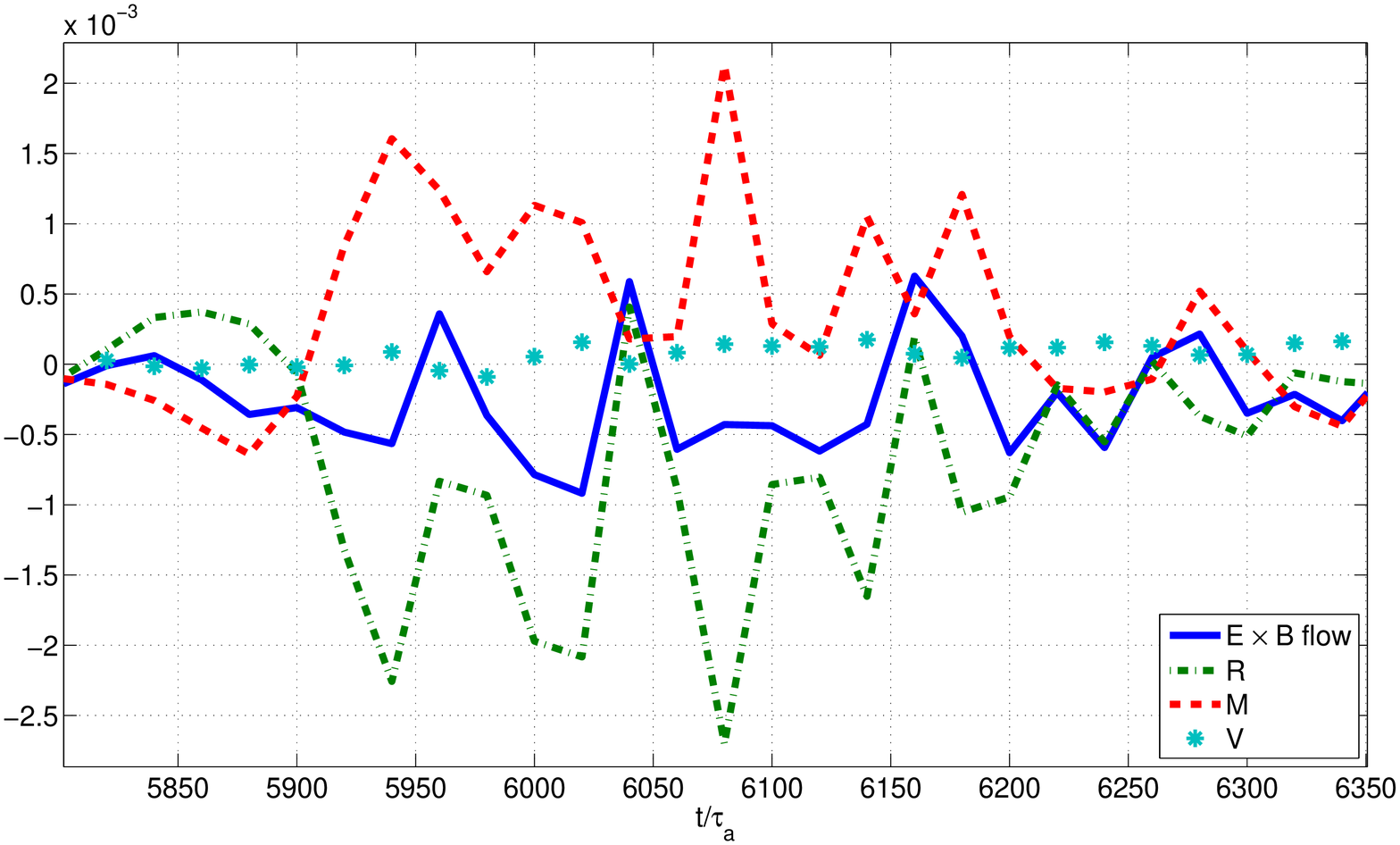}
\end{center}
\caption{Time evolution of M, R and V for $\beta=0.025$ : [Left] Evolution from $t=0\tau_A$ to $t=30000\tau_A$ of the stresses over the current sheet. [Right] Zoom during the transition of the stresses over the current sheet.}
 \label{fig:figure5_c}
\end{figure}

On the left panel of the Figure (\ref{fig:figure5_c}), the time evolutions of R, M and V are presented for a simulation with $\beta=0.025$. As expected, the stresses are nonlinearly generated at the beginning of the first quasi plateau phase allowing the growth of the $E\times B$ poloidal flow. Except at the end of the linear regime where the Reynolds stress is not yet generated, the viscosity term is very weak and does not play an important role in the generation of the flow. The most important contributions come from the Reynolds and Maxwell stresses which balance each other. There is a strong amplification of the amplitude of the stresses during the transition. On the right panel of Figure (\ref{fig:figure5_c}), a closeup of the temporal dynamics of R, M and V during the transition time evolutions are presented. Note that during this transition, whereas the viscosity term is still weak, the evolutions of the Reynolds and Maxwell stresses are complementary. 
%
%
The two stresses tend to balance each other during this phase where the flow is crossing the separatrices, limiting the level of the generated zonal flow, even if they are both growing in amplitude. Once the transition has occured the amplitude of the mean nonlinear brackets in the vicinity of the separatrix, $R$ and $M$, fall. However, in the new dynamical equilibrium the resulting $E\times B$ poloidal flow persists asymptotically and is driven by the Maxwell stress,
as a response of the magnetic structure to the new field distribution.

\subsection{Origin of the nonlinear diamagnetic drift}
In order to investigate the origin of the nonlinear diamagnetic drift, we follow the same procedure for the pressure equation (Eq.\ref{eq:equPE}). After projection on the mode $k_1$, we obtain:
\begin{equation}
\partial_t<\tilde{\omega}^\star>_\delta = dC(t)+dM(t)+D(t)\label{eq:pressure1}
\end{equation}
where
\begin{eqnarray*}
dC(t)&=&-\frac{k_1}{\delta L_y}\int_\delta\int_{L_y}\partial_x[\phi,p]dydx,\\
dM(t)&=&\frac{C^2k_1}{\delta L_y}\int_\delta\int_{L_y}\partial_x[\psi,\nabla_\perp^2\psi]dydx,\\
D(t)&=&\frac{k_1\chi_\perp}{\delta L_y}\int_\delta\int_{L_y}\partial_x^3pdydx.
\end{eqnarray*}

$dC$ is the contribution to the nonlinear diamagnetic drift of the divergence of the convective term, $dM$ is the contribution of the divergence of the Maxwell stress and $D$ is the contribution of the diffusivity.

\begin{figure}
\begin{center}
\includegraphics[width=7cm,height=7cm]{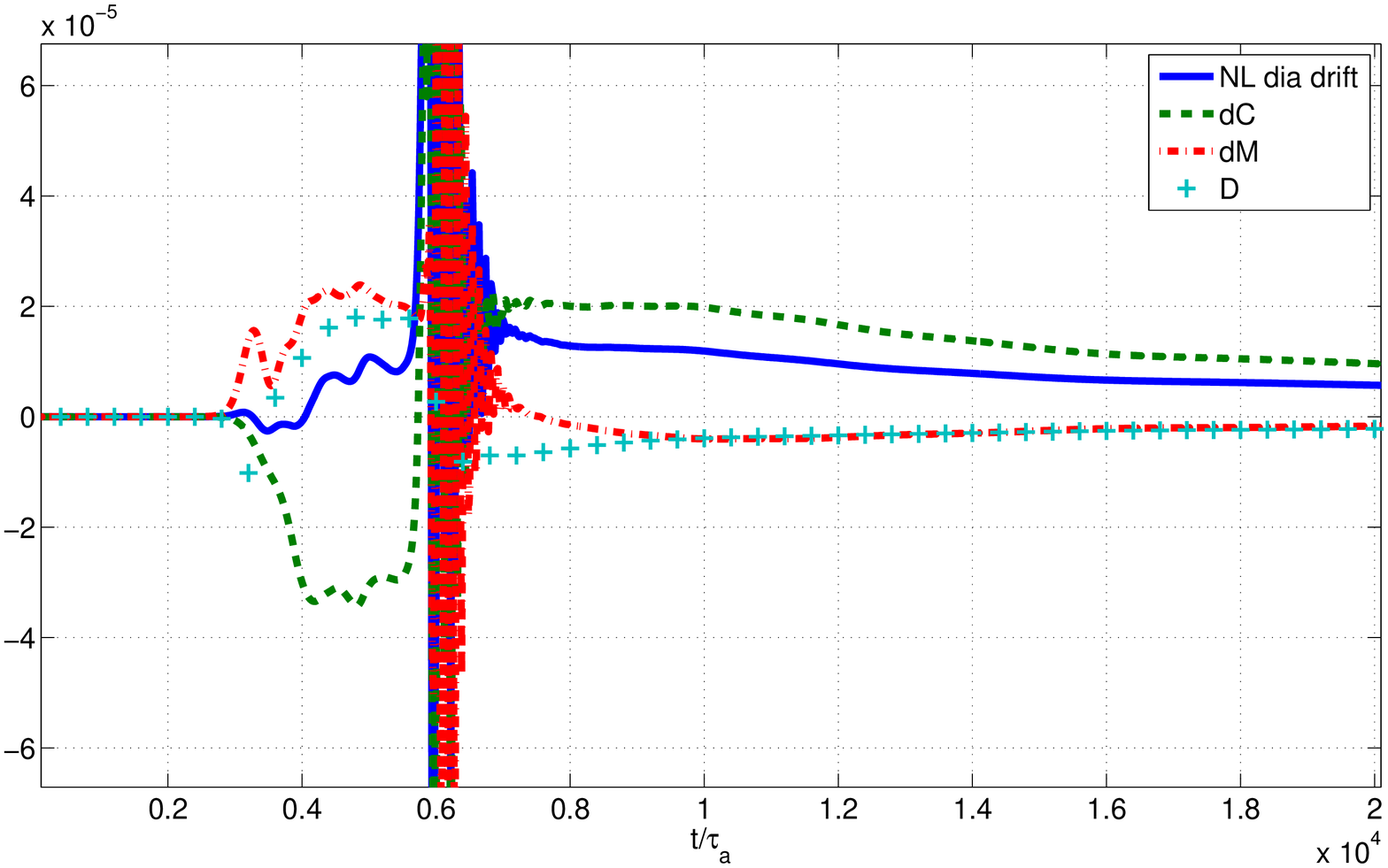}
\includegraphics[width=7cm,height=7cm]{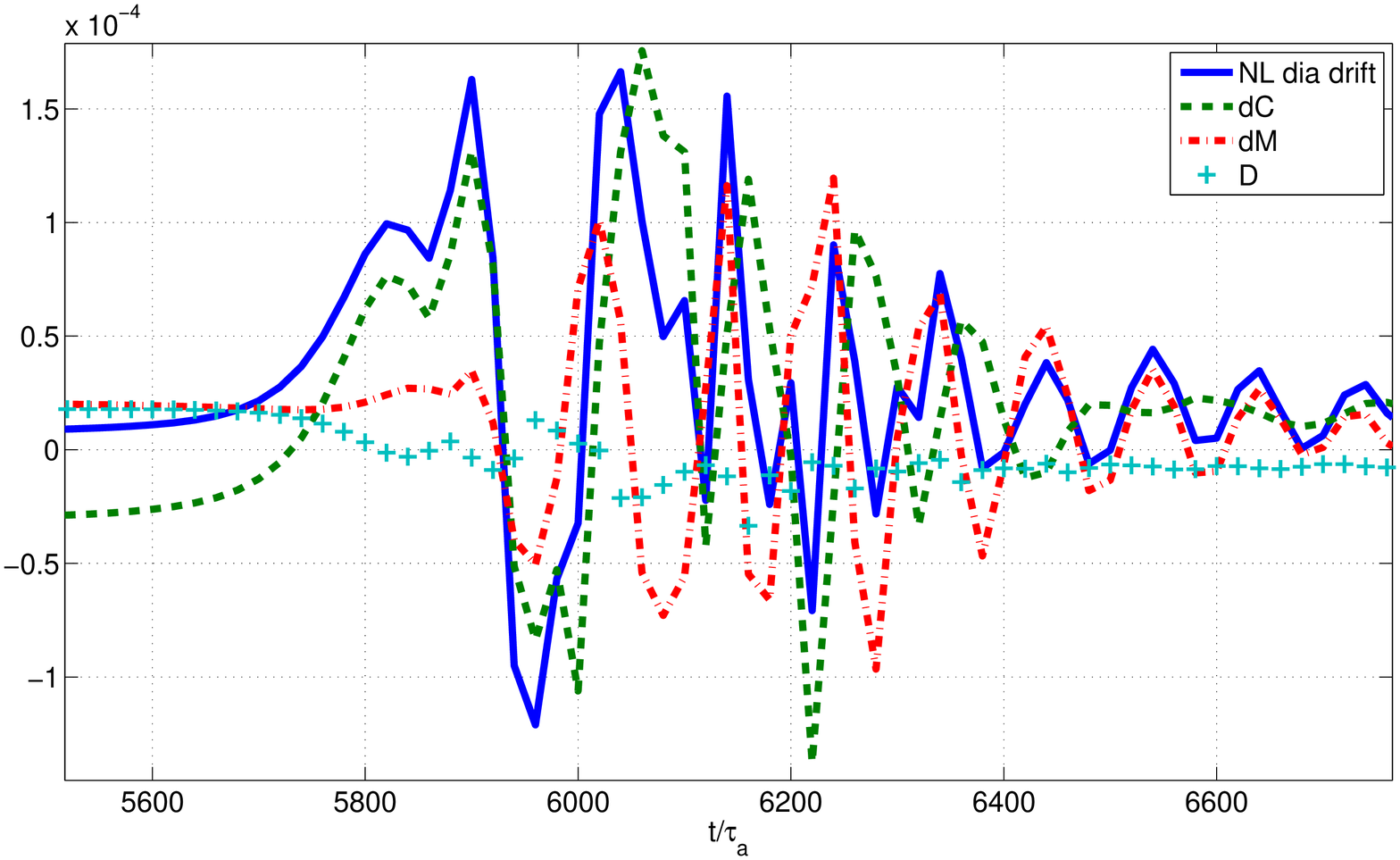}
\end{center}
\caption{Time evolution of dC, dM and D for $\beta=0.025$ : [Left] Evolution from $t=0\tau_A$ to $t=30000\tau_A$ over the current sheet. [Right] Zoom during the tansition over the current sheet.}
 \label{fig:figure5_d}
\end{figure}

On the left panel of Figure (\ref{fig:figure5_d}), the time evolutions of dC, dM and D are presented. The three contributions to the nonlinear diamagnetic drift are generated at the beginning of the nonlinear regime. During the first quasi plateau phase, dC, dM and D participate actively in the generation and growth of the nonlinear diamagnetic drift $\tilde{\omega}^\star$. At the transition there is an amplification of the amplitude of the three contributions. The right panel of Figure (\ref{fig:figure5_d}) presents a closeup of the time evolutions of dC, dM and D during the transition. During this transition, the contribution of the diffusivity does not grow and is relatively weak compared to dC and dM. Note that both the divergence of the convective term and of the Maxwell stress feed the nonlinear diamagnetic drift. 
We observe that in the first phase of the transition, the nonlinear diamagnetic drift is driven by $dC$, the term linked to the advection of the pressure cells. In the following phase this term is balanced by the Maxwell divergence stress leading to a stabilization of the island dynamics. This shows the importance of the coupling parameter $C$ during the transition. In the saturation phase, the origin of $\tilde{\omega}^\star$ comes mainly from the divergence of the convective term, dM and D becoming relatively weak.

\subsection{Effect of the $\beta$ parameter on the island poloidal rotation}
\begin{figure}
\begin{center}
\includegraphics[width=10cm,height=7cm]{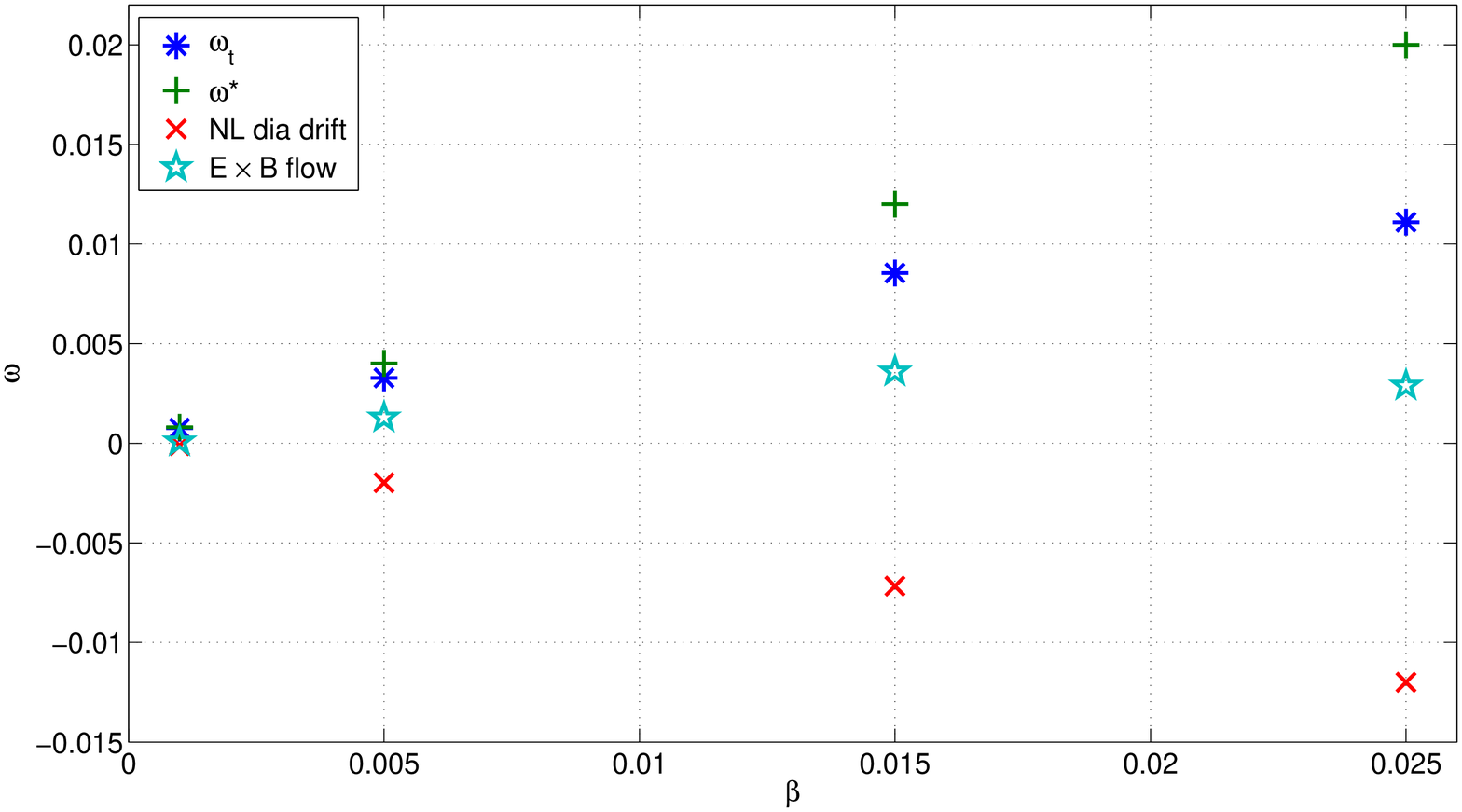}
\end{center}
\caption{Effect of $\beta$ on the rotation frequency after the transition.}
 \label{fig:figure5_e}
\end{figure}

\begin{figure}
\begin{center}
\includegraphics[width=7cm,height=7cm]{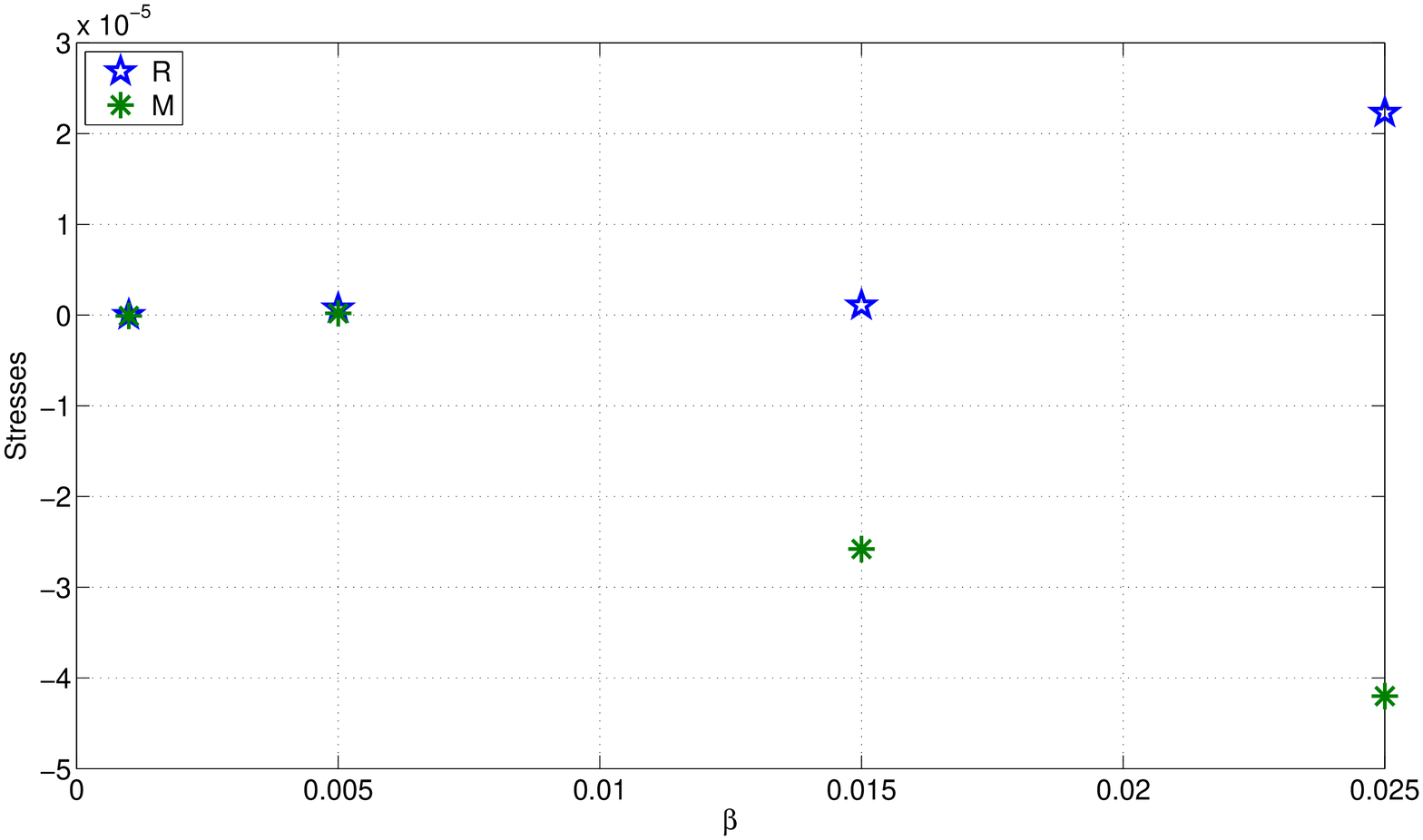}
\includegraphics[width=7cm,height=7cm]{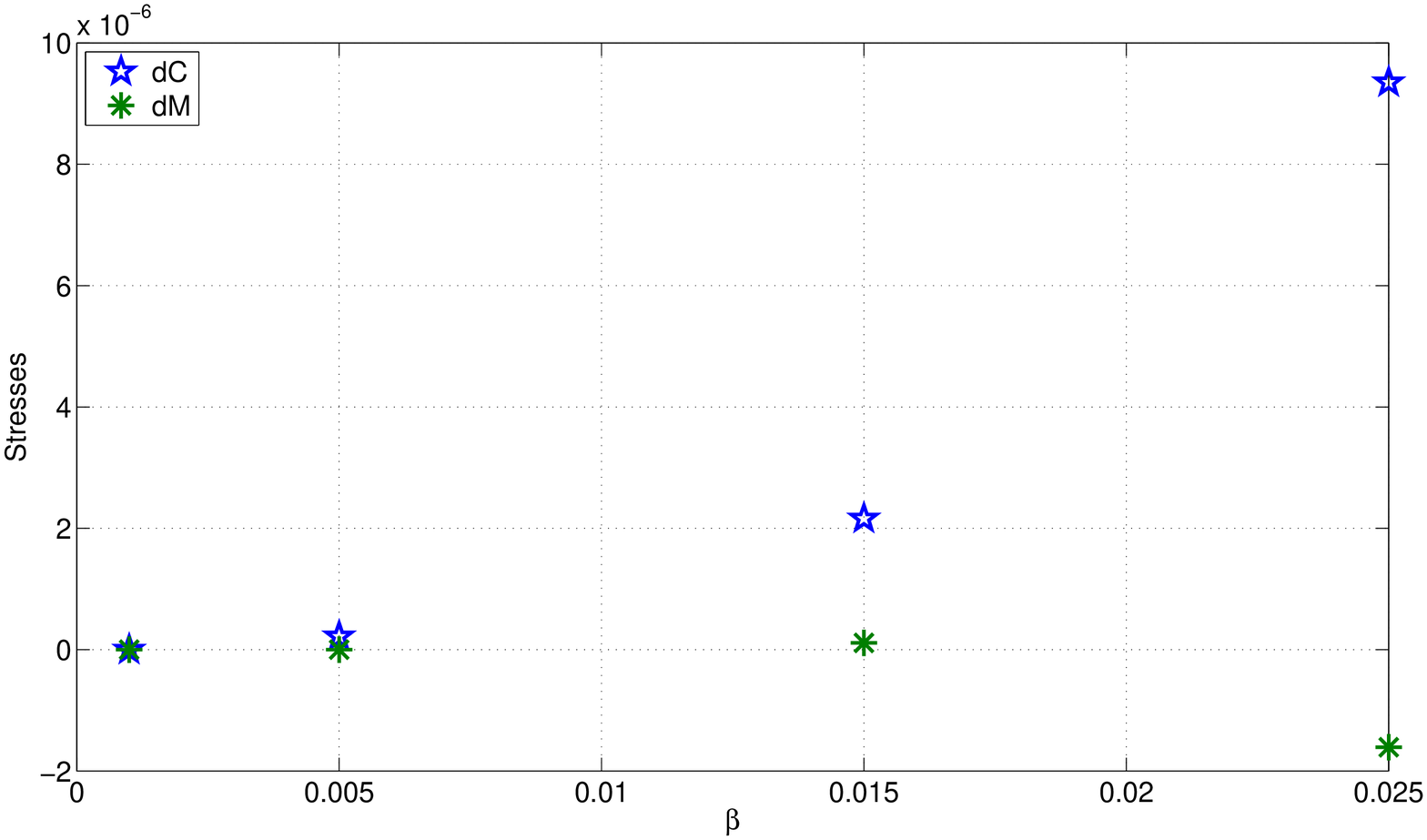}
\end{center}
\caption{Effetc of $\beta$ on the stresses after the transition: [Left] Reynolds and Maxwell stresses versus $\beta$ parameter. [Right] Divergence of the convection and divergence of the Maxwell stress versus $\beta$.}
 \label{fig:figure5_fg}
\end{figure}

In Figure (\ref{fig:figure5_e}), the effect of the $\beta$ parameter on the rotation frequency is presented. The frequencies have been time averaged from $t=15000\tau_A$ to $t=30000\tau_A$. At low $\beta$, pressure effects are weak and the situation is close the classical tearing situation. The competition between Reynolds and Maxwell stresses produces nonlinearly neither the zonal flow nor a diamagnetic drift. The asymptotic rotation frequency $\omega_t$ increases with $\beta$. The nonlinear diamagnetic frequency increases almost linearly with $\beta$, but with a slope lower than the linear one, allowing a global asymptotic drift of the island in the electron diamagnetic direction. Let us note however that the direction of the island rotation depends on the value of the viscosity parameter \cite{Nishimura08}. For any $\beta$, the zonal flow contribution to the island drift is weaker than the diamagnetic one. 
%
%
Note also that from $\beta\sim0.015$, $\omega_{E\times B}$ decreases, and this is linked to the transition observed in Figure (\ref{fig:figure5_fg})
where the effect of the $\beta$ parameter on the average stresses is shown (from $t=15000\tau_A$ to $t=30000\tau_A$). The amplification of the stresses leads to the nonlinear generation of the mean flows and hence affects considerably the rotation of the magnetic island. More precisely, it appears that for low $\beta$ values, the amplitudes of the stresses are very weak as in the classical tearing case. This explains why nonlinear diamagnetic drift and nonlinear $E\times B$ poloidal flow do not affect strongly the rotation of the island in those cases. However, for high $\beta$ regimes, the amplitude of the stresses, in particular the Maxwell and  convective contribution become more important. At $\beta\sim0.015$, a transition is observed : first, $dM$ ceases to be neglectable compared to $dC$, second the Reynolds contribution $R$ starts to grow, weakening the global $E\times B$ flow, as observed in Figure (\ref{fig:figure5_e}).

\section{Summary}
The nonlinear dynamics of a magnetic island in the presence of pressure gradient effects has been investigated. This nonlinear dynamics is different from the classical tearing case and exhibits a bifurcation. After a linear growth of the island and a first quasi plateau phase, a  transition  occurs and the system reaches a new saturated state characterized by the flattening of the pressure profile. We have shown that the dynamics of the island during this bifurcation is due to quasi linear effects. The strong generation of a zonal flow, due to interchange terms, allows this transition to occur. We have shown that the time at which the transition occurs decreases with $\beta$ while it increases with the pressure parameter $\kappa_2$.
Regarding the poloidal rotation of the magnetic island, a model including quasilinear effects has been tested successfully. 
%
Before the transition, the rotation of the island corresponds to the linear diamagnetic drift. Then, at the transition, the rotation is strongly affected by the nonlinear generation of the diamagnetic drift and of the $E\times B$ flow. We have shown that the asymptotic nonlinear diamagnetic drift  is a linear function of $\beta$ but does not cancel the linear drift, as previously obtained when curvature parameters are neglected. The diamagnetic effect appears to be the dominant contribution to the island rotation.  We have shown also that the $\beta$ parameter affects the magnetic rotation through an amplification of the stresses. We have provided a detailed analysis of their impacts on the ${E\times B}$ and diamagnetic drifts for both, the transition and the asymptotic regime. At high $\beta$, we find that a Reynolds stress is generated in the vicinity of the island and weakens the influence of the asymptotic $E\times B$ flow on the rotation.

\ack
The authors want to thank S. Nishimura, D. Escande and G. Fuhr for fruitful discussions. M. M. acknowledges the College Doctoral Franco-Japonais for its fellowship. This work is partly supported by LIA 336 CNRS and by NIFS/NINS under the project of 
Formation of International Network for Scientific Collaborations. This work has been supported by a grant from Agence
  Nationale de la Recherche ANR-05-BLAN-0183-03.\\


\end{document}